\documentclass[12pt, a4paper]{article}
\usepackage{graphicx}
\usepackage{amssymb}
\usepackage{amsmath}
\usepackage{bm}
\usepackage{color}
\usepackage[dvipsnames]{xcolor}
\usepackage{theorem}
\usepackage{subcaption}
\usepackage{listings}
\usepackage{colortbl}
\usepackage{tabularx}
\usepackage{longtable}
\usepackage[utf8]{inputenc}
\usepackage[T1]{fontenc}
\usepackage{lmodern}
\usepackage[top=30truemm,bottom=30truemm,left=25truemm,right=25truemm]{geometry}

\usepackage[sort&compress,numbers, merge]{natbib}

\definecolor{Orange}{cmyk}{0,0.61,0.87,0}
\definecolor{JungleGreen}{cmyk}{0.99,0,0.52,0}
\definecolor{OliveGreen}{cmyk}{0.64,0,0.95,0.40}
\definecolor{Brown}{cmyk}{0,0.81,1,0.60}
\definecolor{RoyalBlue}{cmyk}{0.71,0.53,0,0.12}
\definecolor{Gray}{cmyk}{0,0,0,0.40}
\definecolor{LightPink}{cmyk}{0.0,0.25,0,0}
\definecolor{LLightPink}{cmyk}{0.0,0.10,0,0}
\definecolor{LightBlue}{cmyk}{0.25,0,0,0}
\definecolor{LightGray}{cmyk}{0,0,0,0.2}

\newcommand{\Slash}[1]{{\ooalign{\hfil/\hfil\crcr$#1$}}}

\allowdisplaybreaks[1]

\definecolor{myred}{cmyk}{0,1,1,0.55}
\definecolor{mygreen}{rgb}{0.27, 0.64, 0.48}
\definecolor{mygray}{gray}{.95}

\usepackage[
    colorlinks = true,
    linkcolor = BlueViolet,
    anchorcolor = purple,
    citecolor = magenta,
    filecolor = purple,
    urlcolor = BlueGreen]{hyperref}
\renewcommand{\thefootnote}{\fnsymbol{footnote}}

\begin{document}

\begin{titlepage}

  \begin{flushright}

\end{flushright}

\vskip 2cm
\begin{center}

{\Large 
{\bf
Exploring Chirality Structure in Nucleon Decay
}
}

\vskip 1.5cm

Koichi Hamaguchi$^{a,b}$\footnote{
  E-mail address: \href{mailto:hama@hep-th.phys.s.u-tokyo.ac.jp}{\tt hama@hep-th.phys.s.u-tokyo.ac.jp}}, 
Shihwen Hor$^a$\footnote{
  E-mail address: \href{mailto:shihwen@hep-th.phys.s.u-tokyo.ac.jp}{\tt shihwen@hep-th.phys.s.u-tokyo.ac.jp}},
Natsumi Nagata$^a$\footnote{
E-mail address: \href{mailto:natsumi@hep-th.phys.s.u-tokyo.ac.jp}{\tt natsumi@hep-th.phys.s.u-tokyo.ac.jp}}, 
and 
Hiroki Takahashi$^a$\footnote{
  E-mail address: \href{mailto:takahashi@hep-th.phys.s.u-tokyo.ac.jp}{\tt takahashi@hep-th.phys.s.u-tokyo.ac.jp}}

\vskip 0.8cm

{\it $^a$Department of Physics, University of Tokyo, Bunkyo-ku, Tokyo 113--0033, Japan} \\[2pt]
{\it ${}^b$Kavli Institute for the Physics and Mathematics of the Universe (Kavli IPMU), University of Tokyo, Kashiwa 277--8583, Japan}

\date{\today}

\vskip 1.5cm

\begin{abstract}

Baryon number conservation is an accidental symmetry in the Standard Model, but its violation is theoretically anticipated, making the search for such processes a promising avenue for discovering new physics. In this paper, we explore how measurements of different nucleon decay channels can reveal the structure of the underlying theory. We investigate the chirality structure of baryon-number violating interactions through lifetime measurements of strangeness-conserving nucleon-decay channels. By employing an effective field theory approach, we demonstrate that the ratio of partial decay widths of proton decay channels, $\Gamma(p \to \eta \ell^+)/\Gamma(p \to \pi^0 \ell^+)$, where $\ell^+$ denotes a positron or anti-muon, is sensitive to this chirality structure. Furthermore, we find that in certain new physics models, both anti-lepton and anti-neutrino channels provide valuable insights into the model's structure. Our results highlight the importance of searching for various decay channels in upcoming nucleon decay experiments.

\end{abstract}

\end{center}
\end{titlepage}

\renewcommand{\thefootnote}{\arabic{footnote}}
\setcounter{footnote}{0}

\section{Introduction}

The conservation of baryon number~\cite{Weyl:1929fm, Stueckelberg:1938zz, 41d5269a-4660-3c52-b7c6-7a3f8dc2aff8} is an accidental (classical) symmetry in the Standard Model (SM), but its violation is highly anticipated theoretically. The existence of a matter-antimatter asymmetry in the Universe~\cite{Planck:2018vyg} suggests that baryon-number violating processes were active in the early Universe~\cite{Sakharov:1967dj}. In addition, the unification of gauge interactions and the integration of quarks and leptons in the SM, as predicted by grand unified theories (GUTs), also imply baryon-number violation~\cite{Georgi:1974sy}. Therefore, the search for baryon-number violating processes provides a promising avenue for exploring physics beyond the SM (BSM). Since the 1950s~\cite{Reines:1954pg, Reines:1958pf},\footnote{In addition to direct experimental searches for nucleon decay, theoretical considerations imposed indirect limits on nucleon lifetime based on the observed spontaneous fission rate of Th$^{232}$~\cite{Reines:1954pg} and the heat generation within the Earth~\cite{1959PThPh..22..373Y}.  } nucleon decay has been the focus of experimental searches to investigate baryon-number violation. Despite extensive efforts, it has not yet been observed, resulting in stringent limits on the nucleon decay lifetime. Currently, next-generation nucleon decay experiments with even higher sensitivity, such as JUNO~\cite{JUNO:2021vlw}, Hyper-Kamiokande~\cite{Hyper-Kamiokande:2018ofw}, and DUNE~\cite{DUNE:2020fgq}, are being planned. These experiments are expected to explore lifetimes much longer than the current limits for various nucleon decay channels~\cite{Dev:2022jbf}.

If nucleon decay is observed in future experiments, it would provide immediate evidence of BSM physics, marking it as a significant discovery. However, to fully leverage the potential of these experiments, it is crucial to assess the extent to which we can understand the structure of new physics through the measurement of various nucleon decay channels. In this paper, we study the possibility of exploring the chirality structure of baryon-number violating interactions as comprehensively and generically as possible. Given the stringent existing limits on baryon-number violating processes, it is reasonable to assume that the new-physics effects inducing baryon-number violation sufficiently decouple from the SM. In this case, such effects can be described by dimension-six baryon-number violating effective operators~\cite{Weinberg:1979sa, Wilczek:1979hc, Abbott:1980zj},\footnote{See Ref.~\cite{Beneito:2023xbk} for a recent study on nucleon decays based on effective field theories. } and the Wilson coefficients of these operators contain information about the chirality structure of the underlying physics. As discussed below, this structure influences the nucleon decay branching fractions. Therefore, by observing the various decay channels, we may distinguish potential candidates for the underlying theory.

The outline of this paper is as follows. In Sec.~\ref{sec:eff}, we begin by presenting the dimension-six baryon-number violating effective operators~\cite{Weinberg:1979sa, Wilczek:1979hc, Abbott:1980zj} and demonstrate the matching conditions of their Wilson coefficients above and below the electroweak scale. As our aim is to explore the chirality structure of these effective operators, we focus solely on strangeness-preserving interactions in this paper. In Sec.~\ref{sec:had}, we first summarize the nucleon decay channels considered in this paper (Sec.~\ref{sec:channels}). We then present the relevant hadron matrix elements of the effective operators obtained through QCD lattice simulations using both direct and indirect methods in Sections~\ref{sec:ff} and~\ref{sec:cpt}, respectively. For the subsequent analysis, we utilize the direct method, where the hadron matrix elements are directly computed using lattice simulations. 
Nonetheless, we also discuss the indirect method based on chiral perturbation theory, which clarifies the relationships among nucleon decay channels in terms of $\mathrm{SU}(3)_L \otimes \mathrm{SU}(3)_R$ symmetry. This is useful for understanding the dependence of these decay channels on the Wilson coefficients. In Sec.~\ref{sec:chirality}, we use the effective operators to generically argue that the ratio of the partial decay widths of proton decay channels, $\Gamma (p \to \eta \ell^+)/\Gamma (p \to \pi^0 \ell^+)$ ($\ell^+$ representing a positron or anti-muon), is sensitive to the chirality structure of the effective operators. Channels involving anti-neutrinos also contain information on the chirality structure, but extracting this information is challenging due to the simultaneous contributions from different neutrino final states. However, when considering specific new-physics models, both anti-lepton and anti-neutrino channels can provide valuable insights into the model's structure. We illustrate this in Sec.~\ref{sec:models} with two specific examples, the minimal supersymmetric (SUSY) SU(5) with high-scale SUSY (Sec.~\ref{sec:minimalsu5}) and the minimal SUSY SU(5) with sfermion flavor violation (Sec.~\ref{sec:sfermionfv}). Finally, Sec.~\ref{sec:conclusion} is devoted to our conclusions and discussion.

\section{Effective interactions}
\label{sec:eff}

\subsection{Above the electroweak scale}

Baryon-number violating interactions induced at energy scales much higher than the electroweak scale can be described by non-renormalizable effective operators composed of SM fields. Among these operators, those with the lowest mass dimensions are of dimension six~\cite{Weinberg:1979sa, Wilczek:1979hc, Abbott:1980zj}: 
\begin{align}
 {\cal O}^{(1)}_{ijkl}&=
\epsilon_{abc} \epsilon^{\alpha\beta} (u^a_{Ri}d^b_{Rj})(Q_{Lk\alpha}^c L_{Ll\beta}^{})~,\nonumber \\
 {\cal O}^{(2)}_{ijkl}&=
\epsilon_{abc} \epsilon^{\alpha\beta} ( Q^a_{Li\alpha} Q^b_{Lj\beta})(u_{Rk}^ce_{Rl}^{})~,\nonumber \\
{\cal O}^{(3)}_{ijkl}&=
\epsilon_{abc}\epsilon^{\alpha\beta}\epsilon^{\gamma\delta}
(Q^a_{Li\alpha} Q^b_{Lj\gamma})(Q_{Lk\delta}^c L_{Ll\beta}^{})~,\nonumber \\
 {\cal O}^{(4)}_{ijkl}&=
\epsilon_{abc}(u^a_{Ri}d^b_{Rj})(u_{Rk}^c e_{Rl}^{})~,
\label{eq:fourfermidef}
\end{align}
where $Q_L$, $u_R$, $d_R$, $L_L$, and $e_R$ respectively represent the left-handed quark doublet, the right-handed up-type quark, the right-handed down-type quark, the left-handed lepton doublet, and the right-handed charged lepton fields, all in the two-component notation; $i,j,k,l = 1,2,3$ denote the generations, $a,b,c$ are the color indices, and $\alpha, \beta, \gamma, \delta$ are $\mathrm{SU}(2)_L$ indices; $\epsilon_{abc}$ and $\epsilon^{\alpha\beta}$ are totally anti-symmetric tensors. We denote the Wilson coefficients of these operators by $C^{ijkl}_{(I)}$ for ${\cal O}^{(I)}_{ijkl}$ $(I=1,2,3,4)$:
\begin{align}
  \mathcal{L}_{\mathrm{SM,eff}} = \sum_{I,ijkl} C^{ijkl}_{(I)} {\cal O}^{(I)}_{ijkl} + \mathrm{h.c.} 
  \end{align} 

These operators can be classified into two types. The first two, ${\cal O}^{(1)}_{ijkl}$ and ${\cal O}^{(2)}_{ijkl}$, are composed of both the left- and right-handed fields, which we refer to as the mixed-type operators. The other two consist solely of either left-handed or right-handed fields, and are termed the pure-type operators in this paper. Notice that gauge interactions induce only the mixed-type operators, as each interaction vertex is accompanied by a pair of a Weyl fermion and its conjugate field. Therefore, the presence of the pure-type operators indicates that the underlying physics includes non-gauge interactions. 

The Wilson coefficients $C^{ijkl}_{(I)}$ are determined at the scale of new physics and run down to the electroweak scale using the renormalization group equations (RGEs) given in Refs.~\cite{Abbott:1980zj, Alonso:2014zka}. If there is an energy threshold above which supersymmetry (SUSY) appears, we need to match the Wilson coefficients $C^{ijkl}_{(I)}$ to those of the effective operators defined in SUSY theories, such as the dimension-five superpotential operators and dimension-six K\"{a}hler-potential operators~\cite{Weinberg:1981wj, Sakai:1981pk}. The RGEs for these operators can be found in Refs.~\cite{Munoz:1986kq, Goto:1998qg, Hisano:2013ege}.

\subsection{Below the electroweak scale}

\begin{figure}
  \centering
  \includegraphics[height=70mm]{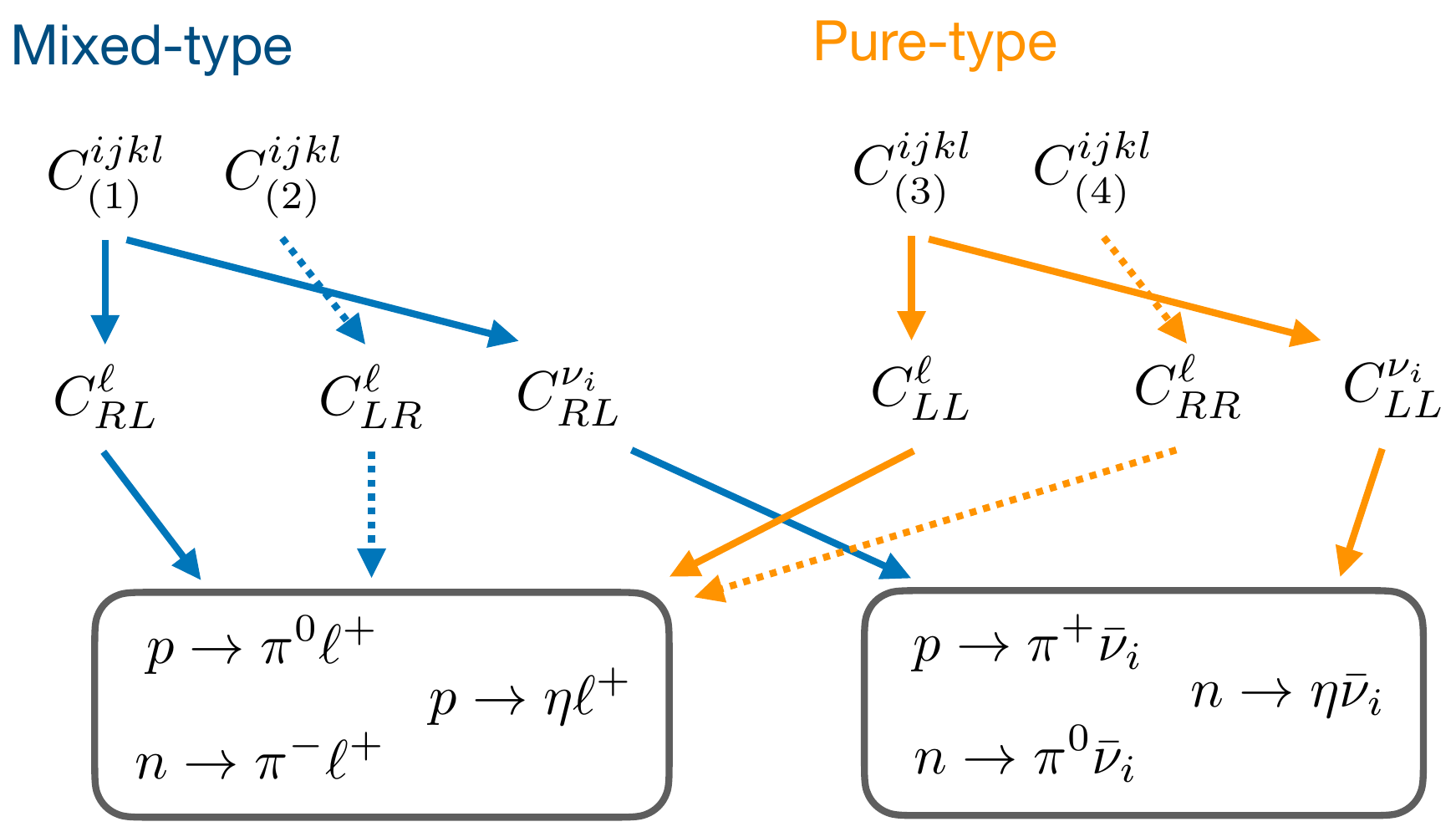}
  \caption{The relations between the Wilson coefficients above and below the electroweak scale, and the nucleon decay channels induced by the corresponding effective interactions. 
  }  
  \label{fig:matching}
\end{figure}

Below the electroweak scale, the baryon-number violating interactions are described by the dimension-six effective operators that preserve strong and electromagnetic interactions. As mentioned in the introduction, in this paper, we focus on the strangeness conserving ($\Delta S = 0$) processes. The relevant effective interactions for these processes are 
\begin{align}
  \mathcal{L}_{\mathrm{eff}} &= C_{RL}^\ell \left[ \epsilon_{abc} (u_R^a d_R^b) (u_L^c \ell_L) \right]
  + C_{LR}^\ell \left[ \epsilon_{abc} (u_L^a d_L^b) (u_R^c \ell_R) \right] \nonumber \\ 
  &+ C_{LL}^\ell \left[ \epsilon_{abc} (u_L^a d_L^b) (u_L^c \ell_L) \right] 
  + C_{RR}^\ell \left[ \epsilon_{abc} (u_R^a d_R^b) (u_R^c \ell_R) \right] \nonumber \\ 
  &+ C_{RL}^{\nu_i} \left[ \epsilon_{abc} (u_R^a d_R^b) (d_L^c \nu_{Li}) \right]
  + C_{LL}^{\nu_i} \left[ \epsilon_{abc} (u_L^a d_L^b) (d_L^c \nu_{Li}) \right] + \mathrm{h.c.} ~,
  \label{eq:leff}
\end{align}
where the quark fields in these operators are defined in the mass eigenstates and $\ell = e, \mu$. The coefficients in Eq.~\eqref{eq:leff} are matched to $C^{ijkl}_{(I)}$ at the electroweak scale as 
\begin{align}
C_{RL}^\ell &= C_{(1)}^{111\ell} ~, \qquad 
C_{LR}^\ell = V_{i1} C^{i11\ell}_{(2)} + V_{j1}  C^{1j1\ell}_{(2)} ~, \nonumber \\ 
C_{LL}^\ell &= - V_{j1} C^{1j1\ell}_{(3)} ~, \qquad 
C_{RR}^\ell = C^{111\ell}_{(4)} ~, \nonumber \\ 
C_{RL}^{\nu_i} &= - V_{j1} C^{11ji}_{(1)} ~, \qquad C_{LL}^{\nu_i} = V_{j1} V_{k1} C^{j1ki}_{(3)} ~,
\label{eq:ewmatching}
\end{align}
where $V$ is the CKM matrix with 
\begin{equation}
  Q_{Li} = 
  \begin{pmatrix}
    u_{Li} \\ V_{ij} d_{Lj}
  \end{pmatrix}
  ~.
\end{equation}
We find that the mixed-type (pure-type) operators at low energies are induced by the mixed-type (pure-type) operators above the electroweak scale, \textit{i.e.}, by $\mathcal{O}^{(1)}_{ijkl}$ and $\mathcal{O}^{(2)}_{ijkl}$ ($\mathcal{O}^{(3)}_{ijkl}$ and $\mathcal{O}^{(4)}_{ijkl}$). In addition, the operators containing the left-handed lepton doublet field, $\mathcal{O}^{(1)}_{ijkl}$ and $\mathcal{O}^{(3)}_{ijkl}$, lead to not only the operators with a charged lepton but also those with a neutrino; the coefficients of these operators differ from each other by the multiplication of the CKM matrix. We illustrate the relations between the Wilson coefficients in Fig.~\ref{fig:matching}. 

The Wilson coefficients in Eq.~\eqref{eq:ewmatching} are evolved down to the hadronic scale with the RGEs given in Ref.~\cite{Nihei:1994tx}, where the corresponding hadron matrix elements are evaluated. We then use them to calculate the nucleon decay rates, as we describe in the subsequent section.

\section{Hadron matrix elements}
\label{sec:had}

\subsection{Nucleon decay channels}
\label{sec:channels}

\begin{table}
  \centering
  \begin{tabular}{lll}
  \hline \hline
   Decay Mode ~~ & Current  [years] ~~ & HK sensitivity  [years]  \\
  \hline
   $p \to \pi^0 e^+ $ & $2.4 \times 10^{34}$~\cite{Super-Kamiokande:2020wjk} & $7.8 \times 10^{34}$~\cite{Hyper-Kamiokande:2018ofw}  \\
   $p \to \pi^0 \mu^+ $ & $1.6 \times 10^{34}$~\cite{Super-Kamiokande:2020wjk} & $7.7 \times 10^{34}$~\cite{Hyper-Kamiokande:2018ofw} \\ 
   $p \to \eta e^+ $ & $1.0 \times 10^{34}$~\cite{Super-Kamiokande:2017gev} & $4.3 \times 10^{34}$~\cite{Hyper-Kamiokande:2018ofw} \\
   $p \to \eta \mu^+ $ & $4.7 \times 10^{33}$~\cite{Super-Kamiokande:2017gev} & $4.9 \times 10^{34}$~\cite{Hyper-Kamiokande:2018ofw} \\
   $p \to \pi^+ \bar{\nu} $ & $3.9 \times 10^{32}$~\cite{Super-Kamiokande:2013rwg} & \\
   \hline 
   $n \to \pi^- e^+ $ & $5.3 \times 10^{33}$~\cite{Super-Kamiokande:2017gev}  & $2.0 \times 10^{34}$~\cite{Hyper-Kamiokande:2018ofw} \\
   $n \to \pi^- \mu^+ $ & $3.5 \times 10^{33} $~\cite{Super-Kamiokande:2017gev}  & $1.8 \times 10^{34}$~\cite{Hyper-Kamiokande:2018ofw} \\
   $n \to \pi^0 \bar{\nu}$ & $1.1 \times 10^{33}$~\cite{Super-Kamiokande:2013rwg}  & \\
   $n \to \eta \bar{\nu} $ & $1.6 \times 10^{32}$~\cite{McGrew:1999nd} & \\
  \hline \hline
  \end{tabular}
  \caption{Nucleon decay channels considered in this work, with the current limits (90\% CL) and future Hyper-Kamiokande sensitivities (90\% CL; 1.9~Megaton$\cdot$year exposure) to their lifetime. 
  }
  \label{tab:channels}
\end{table}

Various nucleon decay channels are induced by the interactions in Eq.~\eqref{eq:leff}. In our analysis, we focus on the two-body $\Delta S = 0$ decay channels that include a pseudo-scalar meson in the final state, summarized in Table~\ref{tab:channels}. As we will discuss in Sec.~\ref{sec:ff}, for these channels, QCD lattice calculations of the hadron matrix elements are available for the operators in Eq.~\eqref{eq:leff}, which allow us to reliably evaluate the partial decay widths of these decay channels. This is in stark contrast to the channels including multiple mesons, a vector meson, or a photon in the final state, for which the precision of theoretical estimates is limited by uncertainties in low-energy constants or assumptions made in the calculation~\cite{Wise:1980ch, Claudson:1981gh, Kaymakcalan:1983uc, Bansal:2022xbg, Fajfer:2023gfi}. 

In Table~\ref{tab:channels}, we also show the current 90\% CL limits on the decay channels~\cite{Super-Kamiokande:2020wjk, Super-Kamiokande:2017gev, Super-Kamiokande:2013rwg, McGrew:1999nd} as well as the expected 90\% CL sensitivities of the Hyper-Kamiokande experiment with 1.9~Megaton$\cdot$year exposure~\cite{Hyper-Kamiokande:2018ofw}.\footnote{The sensitivities achieved by JUNO~\cite{JUNO:2021vlw} and DUNE~\cite{DUNE:2020fgq} are expected to be lower than the Hyper-Kamiokande sensitivities in general; for instance, the DUNE sensitivity for the $p \to \pi^0 e^+$ channel is expected to be $8.7 \times 10^{33}$--$1.1\times 10^{34}$~years, depending on the level of energy smearing.    } 

Notice that the decay widths of the neutron decay channels $n \to \pi^- \ell^+$ are related to those of $p \to \pi^0 \ell^+$ via SU(2) isospin relations as 
\begin{equation}
    \Gamma (n \to \pi^- \ell^+) = 2 \Gamma (p \to \pi^0 \ell^+) ~.
    \label{eq:npieqppi0}
\end{equation}
As we see in Table~\ref{tab:channels}, the experimental sensitivities to these neutron decay channels are weaker than the corresponding proton decay channels by more than a factor of two. We thus consider $p \to \pi^0 \ell^+$ rather than $n \to \pi^- \ell^+$ in what follows. On the other hand, the $n \to \pi^0 \bar{\nu}$ channel is related to the $p \to  \pi^+ \bar{\nu}$ channel by isospin as
\begin{equation}
    \Gamma (n \to \pi^0 \bar{\nu}) = \frac{1}{2} \Gamma (p \to \pi^+ \bar{\nu}) ~. 
    \label{eq:npi0ppi}
\end{equation}
In this case the experimental sensitivity to the neutron decay channel is better than that to the proton decay channel by more than a factor of two. We thus consider $n \to \pi^0 \bar{\nu}$, in stead of $p \to  \pi^+ \bar{\nu}$, in the following analysis.

\subsection{Form factors}
\label{sec:ff}

The quark part of the effective operators in Eq.~\eqref{eq:leff} has the form 
\begin{equation}
  \mathcal{O}_{\chi \chi^\prime} = \epsilon_{abc} (q^a_{\chi} q^{\prime b}_{\chi}) q^{\prime \prime  c}_{\chi^\prime} ~,
  \label{eq:ochichipr}
\end{equation}
where $\chi, \chi^\prime = L/R$ denotes the chirality of the quark fields. The matrix element of this operator between the one-nucleon state $| N (\bm{k})\rangle$ and the one-meson state $|\Pi (\bm{p})\rangle$ can be expressed as\footnote{In the Euclidean spacetime, this is expressed as~\cite{JLQCD:1999dld} 
\begin{equation}
  \langle \Pi (\bm{p}) | \mathcal{O}_{\chi \chi^\prime} | N (\bm{k} ) \rangle = P_{\chi^\prime} \left[ W_0^{\mathcal{O}} (Q^2) - \frac{i\Slash{q}}{m_N} W_1^{\mathcal{O}} (Q^2) \right] u_N (\bm{k})~.
\end{equation}
See Ref.~\cite{Abramczyk:2017oxr} for the conversion between the Minkowski and Euclidean notations. 
} 
\begin{equation}
  \langle \Pi (\bm{p}) | \mathcal{O}_{\chi \chi^\prime} | N (\bm{k} ) \rangle = P_{\chi^\prime} \left[ W_0^{\mathcal{O}} (q^2) + \frac{\Slash{q}}{m_N} W_1^{\mathcal{O}} (q^2) \right] u_N (\bm{k})~,
\end{equation}
where $P_{\chi'}$ denotes the chirality projection operator, $m_N$ is the nucleon mass, $u_N (\bm{k})$ is the four-component nucleon spinor wave function, and $q \equiv k- p$. For the operators in Eq.~\eqref{eq:leff}, the form factors $W_0^{\mathcal{O}}$ and $W_1^{\mathcal{O}}$ are calculated with QCD lattice simulations~\cite{Yoo:2021gql}. With this matrix element, we can calculate the partial decay rate of nucleon. For the decay channels including an anti-lepton in the final state, we have\footnote{For the four-component spinor wave functions, we use the convention $v^c (\bm{q}) \equiv C \bar{v}^T (\bm{q}) = u(\bm{q})$, where $C$ is the charge conjugation matrix. In this case, we have $\overline{v^c}(\bm{q}) (\Slash{q} - m) = \bar{u} (\bm{q}) (\Slash{q} - m) = 0$. Notice that in the Euclidean spacetime this equation has the form $\overline{v^c}(\bm{q}) (i\Slash{q} + m) = \bar{u} (\bm{q}) (i\Slash{q} + m) = 0$ and the first equation here differs from that adopted in Ref.~\cite{Yoo:2021gql} by a sign in front of the fermion mass. This difference results in the opposite sign of $m_\ell$ in Eq.~\eqref{eq:gamnpil}, Eq.~\eqref{eq:wlell}, and Eq.~\eqref{eq:wrell}. 
} 
\begin{align}
  \Gamma (N \to \Pi \ell^+) &= \frac{|\bm{q}_\ell|}{8\pi m_N} \Bigl[E_{\ell}\mathcal{A}_{N\Pi\ell} + {m_\ell} \mathcal{B}_{N\Pi \ell} \Bigr] ~,
  \label{eq:gamnpil}
\end{align}
where $m_\ell$, $\bm{q}_\ell$, and $E_\ell$ are the mass, momentum, and energy of the charged anti-lepton,\footnote{
\begin{align}
  E_\ell &= \frac{m_N^2 -m_{\Pi}^2 + m_{\ell}^2}{2m_N} ~, 
  \qquad 
  |\bm{q}_\ell| = \frac{m_N}{2} \biggl[
  1 - \frac{2 (m_\Pi^2 + m_\ell^2)}{m_N^2} + \frac{(m_\Pi^2 -m_\ell^2)^2}{m_N^4}  
  \biggr]^{\frac{1}{2}} ~,
\end{align}
with $m_\Pi$ the mass of the meson $\Pi$. 
} respectively, and 
\begin{align}
  \mathcal{A}_{N\Pi\ell} &= \left| W_{N\Pi\ell}^L \right|^2 + \left| W_{N\Pi\ell}^R \right|^2 ~, \qquad \mathcal{B}_{N\Pi\ell} = 2 \mathrm{Re} \left[ W_{N\Pi\ell}^L (W_{N\Pi\ell}^R)^* \right]  ~,
\end{align}
with 
\begin{align}
  W_{N\Pi\ell}^L &= C_{RL}^\ell W_{N\Pi\ell, 0}^{RL} (m_\ell^2) + C_{LL}^\ell W_{N\Pi\ell, 0}^{LL} (m_\ell^2) + \frac{m_\ell}{m_N} \Bigl[C_{LR}^\ell W_{N\Pi\ell, 1}^{LR} (m_\ell^2) + C_{RR}^\ell W_{N\Pi\ell, 1}^{RR} (m_\ell^2)\Bigr] ~, \label{eq:wlell} \\ 
  W_{N\Pi\ell}^R &= C_{LR}^\ell W_{N\Pi\ell, 0}^{LR} (m_\ell^2) + C_{RR}^\ell W_{N\Pi\ell, 0}^{RR} (m_\ell^2) + \frac{m_\ell}{m_N} \Bigl[C_{RL}^\ell W_{N\Pi\ell, 1}^{RL} (m_\ell^2) + C_{LL}^\ell W_{N\Pi\ell, 1}^{LL} (m_\ell^2)\Bigr] ~. \label{eq:wrell} 
\end{align}
The values of the matrix elements $W_{N\Pi\ell, 0}^{\chi \chi^\prime} (m_\ell^2)$ and $W_{N\Pi\ell, 1}^{\chi \chi^\prime} (m_\ell^2)$ are given below. 

For the anti-neutrino channels, the expression of the decay width is much simpler than the above expression as we can safely neglect their masses: 
\begin{equation}
  \Gamma (N \to \Pi \bar{\nu}) = \sum_{i = 1,2,3} \frac{m_N}{32\pi} \biggl(1 - \frac{m_\Pi^2}{m_N^2}\biggr)^2 \left|
  C_{RL}^{\nu_i} W_{N\Pi\nu, 0}^{RL} (0) + C_{LL}^{\nu_i} W_{N\Pi\nu, 0}^{LL} (0)
  \right|^2 ~,
  \label{eq:gamnpinu}
\end{equation}
where we take the sum over the neutrino flavors since they are not distinguished in nucleon decay experiments. 

The parity and SU(2) isospin in QCD impose symmetry relations among the hadron matrix elements. The parity invariance of the theory leads to\footnote{Under the parity transformation, we have 
\begin{align}
  \mathcal{O}_{L R} &\to - \gamma^0 \mathcal{O}_{RL} ~, \qquad \mathcal{O}_{L L} \to - \gamma^0 \mathcal{O}_{RR} ~, \nonumber \\  |\Pi \rangle &\to - | \Pi \rangle ~, \qquad |N \rangle \to |N\rangle ~,
\end{align}
where we use the four-component notation for the operator $\mathcal{O}_{\chi \chi^\prime}$ defined in Eq.~\eqref{eq:ochichipr}. } 
\begin{align}
  W^{RL}_{N\Pi\ell, I} = W^{LR}_{N\Pi\ell, I}  ~, \qquad 
  W^{LL}_{N\Pi\ell, I} = W^{RR}_{N\Pi\ell, I} ~,
  \label{eq:parityrel}
\end{align}
for $I = 0,1$. On the other hand, from the SU(2) isospin symmetry, it follows that\footnote{These relations explain Eq.~\eqref{eq:npieqppi0} and Eq.~\eqref{eq:npi0ppi}. } 
\begin{align}
  W^{\chi \chi^\prime}_{p \pi^+ \nu, I} &= \sqrt{2} W^{\chi \chi^\prime}_{p \pi^0 \ell, I} ~, \\ 
  W^{\chi \chi^\prime}_{p \pi^0 \ell, I} &= W^{\chi \chi^\prime}_{n \pi^0 \nu, I} ~, \label{eq:wppinpirel}\\ 
  W^{\chi \chi^\prime}_{p \pi^+ \nu, I} &= W^{\chi \chi^\prime}_{n \pi^- \ell, I} ~, \\ 
  W^{\chi \chi^\prime}_{p \eta \ell, I} &= W^{\chi \chi^\prime}_{n \eta \nu, I} ~. 
  \label{eq:wetarel}
\end{align}
Thanks to these relations, we only need the following eight form factors in our analysis: $W_{p \pi^+ \nu, I}^{LR}$, $W_{p \pi^+ \nu, I}^{LL}$, $W_{p \eta \ell, I}^{LR}$, and $W_{p \eta \ell, I}^{LL}$. 

A recent lattice calculation obtains the values of $W_{p \pi^+ \nu, I}^{LR}$ and $W_{p \pi^+ \nu, I}^{LL}$ for $q^2 = 0$ and $m_\mu^2$ at the renormalization scale of 2~GeV in the $\overline{\mathrm{MS}}$ scheme~\cite{Yoo:2021gql}. The uncertainty of this calculation is estimated to be $\sim 10$\% level; for instance, 
\begin{equation}
  W_{p \pi^+ \nu, 0}^{LL} (0) = 0.151 (14)(8)(26) ~\mathrm{GeV}^2 ~,
  \label{eq:wppinu0}
\end{equation}
where the first, second, and third parentheses correspond to the statistical uncertainty, systematic uncertainty due to excited states, and systematic uncertainty due to the continuum extrapolation, respectively. This indicates that $\mathcal{O} (m_\ell / m_N)$ corrections in Eq.~\eqref{eq:gamnpil} can be relevant only for anti-muon for the current level of accuracy in the calculation. We thus keep only the $\mathcal{O} (m_\mu / m_N)$ corrections in the following analysis. It is also found that the difference of the form factors between $q^2 = 0$ and $m_\mu^2$ is as small as $\sim 1$\%, and thus negligible in the present study. For example,  
\begin{equation}
  W_{p \pi^+ \nu, 0}^{LL} (m_\mu^2) = 0.153 (14)(7)(26) ~\mathrm{GeV}^2 ~,
\end{equation}
which is fairly close to the $q^2 = 0$ value in Eq.~\eqref{eq:wppinu0}.  

The form factors $W_{p \eta \ell, 0}^{LR}$ and $W_{p \eta \ell, 0}^{LL}$ are computed in Ref.~\cite{Aoki:2017puj}. It is found that the uncertainty of the pure-type form factor $W_{p \eta \ell, 0}^{LL}$ is comparable to that for the pion channels, but the mixed-type one $W_{p \eta \ell, 0}^{LR}$ suffers from a rather large relative uncertainty ($\sim 50$\%) due to its small central value. On the other hand, $W_{p \eta \ell, 1}^{LR}$ and $W_{p \eta \ell, 1}^{LL}$ are not directly calculated, but the following combinations of the form factors are obtained~\cite{Aoki:2017puj}: 
\begin{equation}
  W_{p \eta \ell, \mu}^{\chi \chi^\prime} \equiv W_{p \eta \ell, 0}^{\chi \chi^\prime} + \frac{m_\mu}{m_N} W_{p \eta \ell, 1}^{\chi \chi^\prime} ~.
  \label{eq:wmudef}
\end{equation}
In the following analysis, we extract $W_{p \eta \ell, 1}^{LR}$ and $W_{p \eta \ell, 1}^{LL}$ from these quantities by using the values of $W_{p \eta \ell, 0}^{LR}$ and $W_{p \eta \ell, 0}^{LL}$.\footnote{We neglect the $q^2$ dependence of these quantities in this prescription.} Throughout this work, we use the values of particle masses given in the Review of Particle Physics~\cite{ParticleDataGroup:2024cfk} unless otherwise noted.  

It is important to notice that the simulation in Ref.~\cite{Aoki:2017puj} does not take into account the contribution of disconnected diagrams to these matrix elements, which may cause additional uncertainty. All in all, the current accuracy of the calculation for the $\eta$ channels may be worse than the $\sim 10$\% level---we anticipate this will be improved with the help of future lattice calculations. 

\begin{table}
  \centering
  \begin{tabular}{c|cc|cc}
  \hline \hline
  $I$ &\multicolumn{2}{c|}{0}&\multicolumn{2}{c}{1}\\
  \hline 
   $\chi \chi^\prime$ & LR & LL  & LR & LL  \\
  \hline
   $W_{p \pi^+ \nu}$, $\sqrt{2} W_{p \pi^0 \ell}$, $\sqrt{2} W_{n \pi^0 \nu}$, $W_{n \pi^- \ell}$ & $-0.159$ & $0.151$ & $0.169$ & $-0.134$ \\
   $W_{p \eta \ell}$, $W_{n \eta \nu}$ & $0.006$ & $0.113$ & $0.044$ & $-0.044$ \\
  \hline \hline
  \end{tabular}
  \caption{The central values of the form factors in units of $\mathrm{GeV}^2$, evaluated at the renormalization scale of 2~GeV in the $\overline{\mathrm{MS}}$ scheme. We use the results given in Ref.~\cite{Yoo:2021gql} and Ref.~\cite{Aoki:2017puj} for the pion and $\eta$ channels, respectively. The values of $W_{p \eta \ell, 1}^{\chi \chi^\prime}$ and $W_{n\eta \nu, 1}^{\chi \chi^\prime}$ are obtained from the results in Ref.~\cite{Aoki:2017puj} according to the prescription described in the text. 
  }
  \label{tab:formfactors}
\end{table}

We summarize the central values of the form factors considered in this paper in Table~\ref{tab:formfactors}.

\subsection{Chiral perturbation theory}
\label{sec:cpt}

Another method for calculating hadron matrix elements is to use baryon chiral perturbation theory~\cite{Claudson:1981gh, Kaymakcalan:1983uc}. QCD lattice simulations are also valuable for this approach, as they can determine the low-energy constants of the theory, thereby significantly enhancing the precision of the calculations. This combined use of lattice computation and chiral perturbation theory is known as the indirect method. This contrasts with the direct method discussed in Sec.~\ref{sec:ff}, where hadron matrix elements are directly computed using lattice simulations. In the indirect method, the expressions for the matrix elements incorporate the flavor $\mathrm{SU}(3)_L \otimes \mathrm{SU}(3)_R$ symmetry structure of QCD due to the use of chiral perturbation theory. This allows for consideration of the relationships among the hadron matrix elements in addition to Eqs.~(\ref{eq:parityrel}--\ref{eq:wetarel}), which helps systematically understand the dependence of nucleon-decay branching fractions on the Wilson coefficients. Therefore, in the following discussion, we will also consider hadron matrix elements evaluated using the indirect method, even though we use those obtained with the direct method for our numerical computations.

At the leading order in the chiral perturbation, the form factors considered in Sec.~\ref{sec:ff} are expressed as follows~\cite{JLQCD:1999dld}: 
\begin{align}
  W_{p \pi^+ \nu, 0}^{LR} &=  \frac{1 + D + F}{ f} \alpha ~, \qquad 
  W_{p \pi^+ \nu, 1}^{LR} = - \frac{2(D + F)}{ f} \alpha ~, \nonumber \\ 
  W_{p \pi^+ \nu, 0}^{LL} &=  \frac{1 + D + F}{ f} \beta ~, \qquad 
  W_{p \pi^+ \nu, 1}^{LL} = - \frac{2(D + F)}{ f} \beta ~, \nonumber \\ 
  W_{p \eta \ell, 0}^{LR} &=  -\frac{1 + D -3 F}{\sqrt{6} f} \alpha ~, \qquad 
  W_{p \eta \ell, 1}^{LR} =  \frac{2(D -3 F)}{\sqrt{6} f} \alpha ~, \nonumber \\ 
  W_{p \eta \ell, 0}^{LL} &=  \frac{3 - D +3 F}{\sqrt{6} f} \beta ~, \qquad 
  W_{p \eta \ell, 1}^{LL} =  \frac{2(D -3 F)}{\sqrt{6} f} \beta ~, 
  \label{eq:ffind}
\end{align}
where $f = 130.2(1.2)$~MeV~\cite{ParticleDataGroup:2024cfk} is the pion decay constant, $D = 0.8$, and $F = 0.47$~\cite{Yoo:2021gql}. The low-energy constants $\alpha$ and $\beta$ are defined by 
\begin{align}
  \langle 0| \epsilon_{abc} (u_R^a d_R^b) u_L^c |p\rangle &\equiv  \alpha P_L u_p ~, \nonumber \\ 
  \langle 0| \epsilon_{abc} (u_L^a d_L^b) u_L^c |p\rangle &\equiv  \beta P_L u_p ~, 
\end{align}
and their values are computed in Ref.~\cite{Yoo:2021gql} as 
\begin{align}
  \alpha &= -0.01257 (111)~\mathrm{GeV}^3 ~, \nonumber \\ 
  \beta &= 0.01269 (107)~\mathrm{GeV}^3 ~.
\end{align}
As we see, these low-energy constants are determined with the precision of $\sim 10$\%. It is also found that to very good precision ($\sim 1$\%), we have\footnote{This implies that the overlap of the one-particle state of proton with the interpolating field $\epsilon_{abc} (\overline{u^C}{}^ad^b) \gamma_5 u^c $ is very small.  } 
\begin{equation}
  \alpha \simeq - \beta ~.
  \label{eq:alpeqmibeta}
\end{equation}
This observation suggests that the uncertainty from the hadron matrix elements is significantly reduced when considering the ratio of partial decay widths, as the dependence on the low-energy constants $\alpha$ and $\beta$, as well as the pion decay constant $f$, is cancelled out. It will be intriguing to verify this expectation using the direct method in future lattice calculations.

\begin{table}
  \centering
  \begin{tabular}{c|cc|cc}
  \hline \hline
  $I$ &\multicolumn{2}{c|}{0}&\multicolumn{2}{c}{1}\\
  \hline 
   $\chi \chi^\prime$ & LR & LL  & LR & LL  \\
  \hline
   $W_{p \pi^+ \nu}$, $\sqrt{2} W_{p \pi^0 \ell}$, $\sqrt{2} W_{n \pi^0 \nu}$, $W_{n \pi^- \ell}$ & $-0.22$ & $0.22$ & $0.25$ & $-0.25$ \\
   $W_{p \eta \ell}$, $W_{n \eta \nu}$ & $0.015$ & $0.14$ & $0.048$ & $-0.049$ \\
  \hline \hline
  \end{tabular}
  \caption{The central values of the form factors in units of $\mathrm{GeV}^2$, evaluated with the indirect method. 
  }
  \label{tab:ind}
\end{table}

In Table~\ref{tab:ind}, we show the central values of the form factors obtained with the indirect method. Comparing these values with those in Table~\ref{tab:formfactors}, we see that the indirect method is able to reproduce the results obtained with the direct method qualitatively, though there are $\mathcal{O}(10)$\% discrepancies numerically.

\section{Chirality structure and nucleon decay branch}
\label{sec:chirality}

We now study the dependence of the nucleon decay widths on the Wilson coefficients of the effective operators in Eq.~\eqref{eq:leff}. Let us first consider the limiting cases where only either the mixed-type or pure-type operators are present. We also take $m_\ell = 0$ in this discussion. 

In the case where only the mixed-type operators are present, \textit{i.e.}, for $C^\ell_{LL} = C^\ell_{RR} = C^{\nu_i}_{LL} = 0$, the nucleon decay widths, Eq.~\eqref{eq:gamnpil} and Eq.~\eqref{eq:gamnpinu}, are given by 
\begin{align}
  \Gamma (N \to \Pi \ell^+) &= \frac{m_N}{32\pi} \biggl(1 - \frac{m_\Pi^2}{m_N^2}\biggr)^2 |W^{LR}_{N\Pi \ell, 0}|^2 \left[ |C_{RL}^\ell|^2 + |C_{LR}^\ell|^2 \right]  ~, \\ 
  \Gamma (N \to \Pi \bar{\nu}) &= \frac{m_N}{32\pi} \biggl(1 - \frac{m_\Pi^2}{m_N^2}\biggr)^2 \left|W_{N\Pi\nu, 0}^{RL} \right|^2 \sum_i \left|
  C_{RL}^{\nu_i}  \right|^2 ~,
\end{align}
where we have used the relation~\eqref{eq:parityrel}. We observe that the decay widths for the charged lepton (neutrino) channels are all proportional to the same factor, $|C_{RL}^\ell|^2 + |C_{LR}^\ell|^2$ ($\sum_i \left|C_{RL}^{\nu_i}  \right|^2$). Notably, the ratios of the decay widths in the charged lepton and neutrino channels, $\Gamma (p \to \eta e^+)/\Gamma (p \to \pi^0 e^+)$ and $\Gamma (n \to \eta \bar{\nu})/\Gamma (n \to \pi^0  \bar{\nu})$, become
independent of the Wilson coefficients and are determined solely by low-energy quantities. It is also found that this conclusion holds true in the case where only the pure-type operators are present, \textit{i.e.}, when $C^\ell_{LR} = C^\ell_{RL} = C^{\nu_i}_{RL} = 0$, although the resultant ratios are different from those in the previous case. We will determine these ratios in this section and demonstrate that they can be used to distinguish these two cases. 

As we see below, the above conclusion is avoided if both the mixed- and pure-type operators exist simultaneously, or if $m_\ell \neq 0$. The ratios of the decay widths between the charged lepton and neutrino channels also depend on the Wilson coefficients. We will see that such dependence is useful to explore the structure of the theory that generates the effective interactions in Eq.~\eqref{eq:leff}. 

In the following analysis, we focus on the ratios of the decay widths rather than the decay widths themselves. This approach is particularly useful when all the Wilson coefficients are generated at the same mass scale, as is the case in GUTs. In such scenarios, the mass scale and a significant portion of the renormalization factors cancel out in the ratios. Furthermore, as discussed in Sec.~\ref{sec:cpt}, we anticipate that the uncertainty from the hadron matrix elements will be reduced in the ratios, although this expectation needs to be quantitatively verified through lattice simulations.

\subsection{Charged anti-lepton channels}

In this subsection, we consider the ratios $\Gamma (p \to \eta \ell^+)/\Gamma (p \to \pi^0 \ell^+)$ ($\ell = e, \mu$). As shown in Fig.~\ref{fig:matching}, all of the operators in Eq.~\eqref{eq:fourfermidef} contribute to these ratios. 

\subsubsection{Positron channels}
\label{sec:positron}

First, let us study the case of $\ell = e$, for which we can take $m_\ell = 0$. As discussed above, in this case, the ratio is independent of the Wilson coefficients if only either the mixed-type or pure-type operators are present: 
\begin{align}
  \frac{\Gamma (p \to \eta e^+)}{\Gamma (p \to \pi^0  e^+)} = \frac{(1-m_\eta^2/m_p^2)^2}{(1-m_\pi^2/m_p^2)^2} \cdot \frac{|W_{p\eta \ell}^{\chi\chi^\prime}|^2}{|W_{p\pi^0 \ell}^{\chi\chi^\prime}|^2}
  \simeq 
  \begin{cases}
    0.0013 & \text{(Mixed only)} \\ 
    0.51 & \text{(Pure only)} 
  \end{cases}
  ~.
  \label{eq:poslim}
\end{align}
The difference between these two values is significant enough to distinguish the two cases, even when considering the $\sim 10$\% error in hadron matrix elements. 

Things get more complicated if there are both the mixed- and pure-type operators. To see the dependence of the ratio on the Wilson coefficients, we express the decay widths using the form factors in Eq.~\eqref{eq:ffind}: 
\begin{align}
  \Gamma (p \to \pi^0 e^+) &= \frac{m_p}{32\pi} \biggl(1-\frac{m_\pi^2}{m_p^2}\biggr)^2 \frac{(1+D+F)^2 \alpha^2}{2f^2} \left[\left|C_{RL}^e - C_{LL}^e\right|^2 + \left|C_{LR}^e - C_{RR}^e\right|^2\right] ~, \\ 
  \Gamma (p \to \eta e^+) &= \frac{m_p}{32\pi} \biggl(1-\frac{m_\eta^2}{m_p^2}\biggr)^2 \frac{\alpha^2}{6 f^2} 
  \Bigl[\left|(1+D-3F)C_{RL}^e +(3-D+3F) C_{LL}^e\right|^2 
  \nonumber \\ 
  &\qquad + \left|(1+D-3F)C_{LR}^e +(3-D+3F) C_{RR}^e\right|^2\Bigr] ~,
\end{align}
where we have used Eq.~\eqref{eq:alpeqmibeta} to eliminate $\beta$ from these expressions. As we see, these decay widths have different dependence on the Wilson coefficients. In particular, the $p \to \pi^0 e^+$ channel is suppressed when 
\begin{equation}
  \frac{C_{LL}^e}{C_{RL}^e} = \frac{C_{RR}^e}{C_{LR}^e} = 1 ~,
\end{equation}
while for the $p \to \eta e^+$ channel the suppression occurs when 
\begin{equation}
  \frac{C_{LL}^e}{C_{RL}^e} = \frac{C_{RR}^e}{C_{LR}^e} = -\frac{1+D-3F}{3-D+3F} \simeq -0.1 ~.
  \label{eq:etamodesup}
\end{equation}
This observation indicates that the determination of the ratio between these decay widths can constrain a linear combination of the ratios of these Wilson coefficients, say, $C_{LL}^e/C^e_{RL}$, $C_{RR}^e/C^e_{RL}$, and $C_{LR}^e/C^e_{RL}$.

\begin{figure}
  \centering
  \includegraphics[height=80mm]{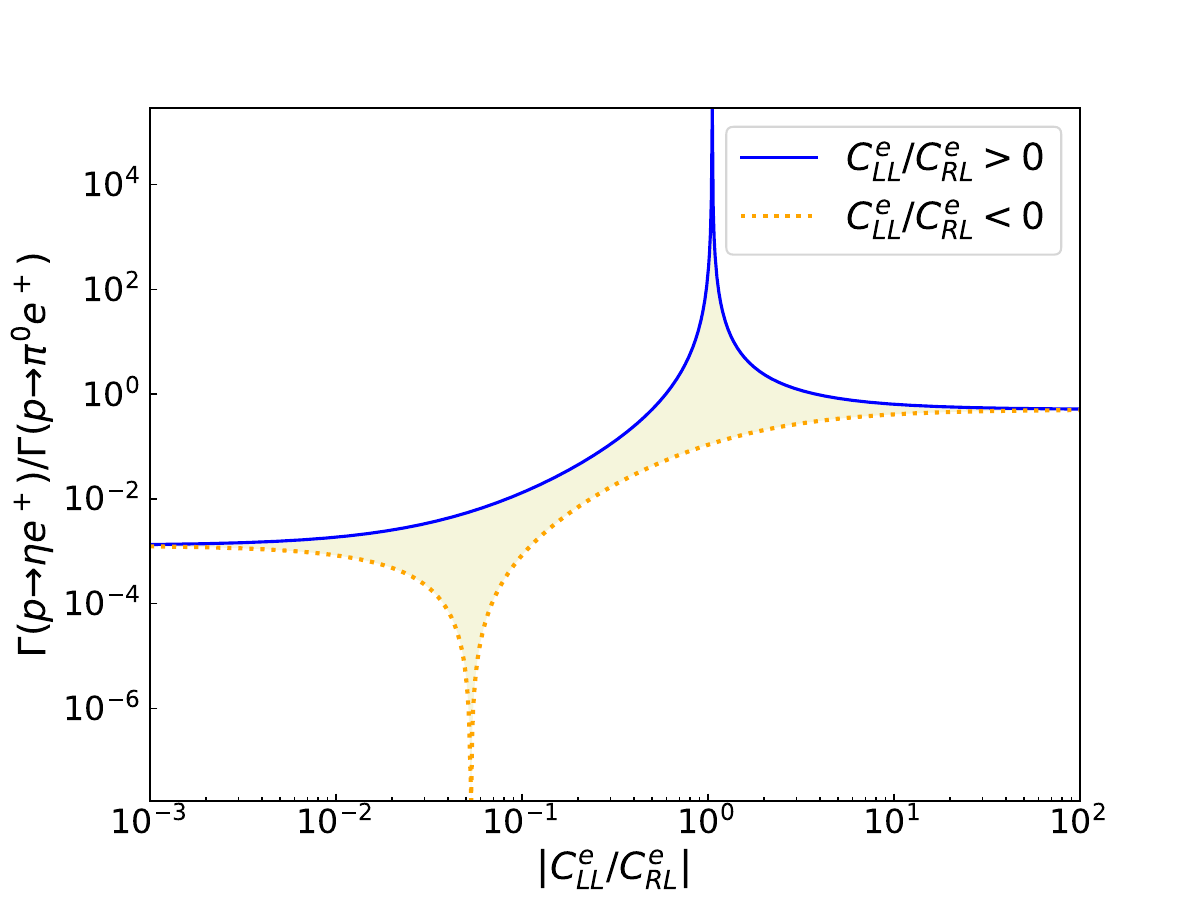}
  \caption{The ratio $\Gamma (p \to \eta e^+)/\Gamma (p \to \pi^0 e^+)$ as a function of $|C_{LL}^e/C_{RL}^e|$ with $C_{RR}^e = C_{LR}^e = 0$. The blue solid and orange dotted lines correspond to $C_{LL}^e/ C_{RL}^e > 0$ and $C_{LL}^e / C_{RL}^e < 0$, respectively, while the shaded region represents the cases where $C_{LL}^e/C^e_{RL}$ is complex.  
  }  
  \label{fig:ratioegeneric}
\end{figure}

For example, in Fig.~\ref{fig:ratioegeneric}, we show the ratio $\Gamma (p \to \eta e^+)/\Gamma (p \to \pi^0 e^+)$ as a function of $|C_{LL}^e/C_{RL}^e|$ with $C_{RR}^e = C_{LR}^e = 0$. The blue solid and orange dotted lines correspond to $C_{LL}^e/ C_{RL}^e > 0$ and $C_{LL}^e / C_{RL}^e < 0$, respectively, while the shaded region represents the cases where $C_{LL}^e/C^e_{RL}$ is complex. We see that the ratio converges to the values in Eq.~\eqref{eq:poslim} for $|C_{LL}^e/C_{RL}^e| \ll 1$ and $|C_{LL}^e/C_{RL}^e| \gg 1$, as they correspond to the mixed-only and pure-only cases, respectively. For $C_{LL}^e/ C_{RL}^e > 0$, the enhancement of the ratio can be found at $C_{LL}^e \simeq C_{RL}^e $ due to the suppression in the $p \to \pi^0 e^+$ channel discussed above. On the other hand, for $C_{LL}^e/ C_{RL}^e < 0$, the point of the suppression in the ratio somewhat deviates from Eq.~\eqref{eq:etamodesup}; this is due to the difference in the hadron matrix elements obtained with the direct and indirect methods, as can be seen in Table~\ref{tab:formfactors} and Table~\ref{tab:ind}.

\subsubsection{Anti-muon channels}

Next, we consider the anti-muon channels, for which we take the effect of the muon mass $m_\mu$ into account. We will show that in this case the ratio $\Gamma (p \to \eta \mu^+)/\Gamma (p \to \pi^0 \mu^+)$ can depend on the Wilson coefficients even when only either the mixed-type or pure-type operators are present.

For the mixed-only case, where $C_{LL}^\mu = C_{RR}^\mu = 0$, we have 
\begin{align}
  \Gamma (p \to \Pi \mu^+) &= \frac{m_p}{32\pi} \biggl(1 - \frac{m_\Pi^2}{m_p^2}\biggr)^2 \left( W^{LR}_{p\Pi \ell, 0} \right)^2 \left[ |C_{RL}^\mu|^2 + |C_{LR}^\mu|^2 \right]  ~, \nonumber\\ 
  &\times \biggl[1 + \frac{4 m_\mu}{m_p} \biggl\{ \frac{W^{LR}_{p\Pi \ell, 1}}{W^{LR}_{p\Pi \ell, 0}} + \biggl(1 - \frac{m_\Pi^2}{m_p^2}\biggr)^{-1} \biggr\} \frac{\mathrm{Re}(C_{RL}^{\mu} C_{LR}^{\mu *})}{|C_{RL}^\mu|^2 + |C_{LR}^\mu|^2}\biggr] ~, 
  \label{eqp:ppimuappr}
\end{align}
while for the pure-only case with $C_{LR}^\mu = C_{RL}^\mu = 0$, 
\begin{align}
  \Gamma (p \to \Pi \mu^+) &= \frac{m_p}{32\pi} \biggl(1 - \frac{m_\Pi^2}{m_p^2}\biggr)^2 \left( W^{LL}_{p\Pi \ell, 0} \right)^2 \left[ |C_{LL}^\mu|^2 + |C_{RR}^\mu|^2 \right]  ~, \nonumber\\ 
  &\times \biggl[1 + \frac{4 m_\mu}{m_p} \biggl\{ \frac{W^{LL}_{p\Pi \ell, 1}}{W^{LL}_{p\Pi \ell, 0}} + \biggl(1 - \frac{m_\Pi^2}{m_p^2}\biggr)^{-1} \biggr\} \frac{\mathrm{Re}(C_{LL}^{\mu} C_{RR}^{\mu *})}{|C_{LL}^\mu|^2 + |C_{RR}^\mu|^2}\biggr] ~.
\end{align}
We have retained terms up to the order of $\mathcal{O}(m_\mu/m_p)$ in these expressions. The ratio $\Gamma(p \to \eta \mu^+)/\Gamma(p \to \pi^0 \mu^+)$ is found to depend on the Wilson coefficients due to these $\mathcal{O}(m_\mu/m_p)$ terms. The terms are maximized when $|C_{RL}^\mu| \simeq |C_{LR}^\mu|$ in the mixed-only case, and $|C_{LL}^\mu| \simeq |C_{RR}^\mu|$ in the pure-only case. If one coefficient significantly dominates the other, the $m_\mu$-dependent term is suppressed, leading to the same results as in Eq.~\eqref{eq:poslim}.  It is also found that the dependence on the Wilson coefficients is enhanced when $|W^{\chi \chi^\prime}_{p\Pi \ell, 1}| \gg |W^{\chi \chi^\prime}_{p\Pi \ell, 0}|$. Table~\ref{tab:formfactors} shows that the ratio $|W^{\chi \chi^\prime}_{p\Pi \ell, 1}| / |W^{\chi \chi^\prime}_{p\Pi \ell, 0}|$ is maximized for the mixed-type operators and $\Pi = \eta$. Therefore, we expect the dependence of the ratio $\Gamma(p \to \eta \mu^+)/\Gamma(p \to \pi^0 \mu^+)$ on the Wilson coefficients to be more pronounced in the mixed-only case than in the pure-only case. However, for the mixed-type operators, the form factor $W^{LR}_{p\eta \ell, 0}$ is numerically small, resulting in a smaller value for the ratio, as previously discussed in Sec.~\ref{sec:positron}.

\begin{figure}
      \centering
      \subcaptionbox{\label{fig:mumix}
      Mixed only
      }
      {\includegraphics[width=0.48\textwidth]{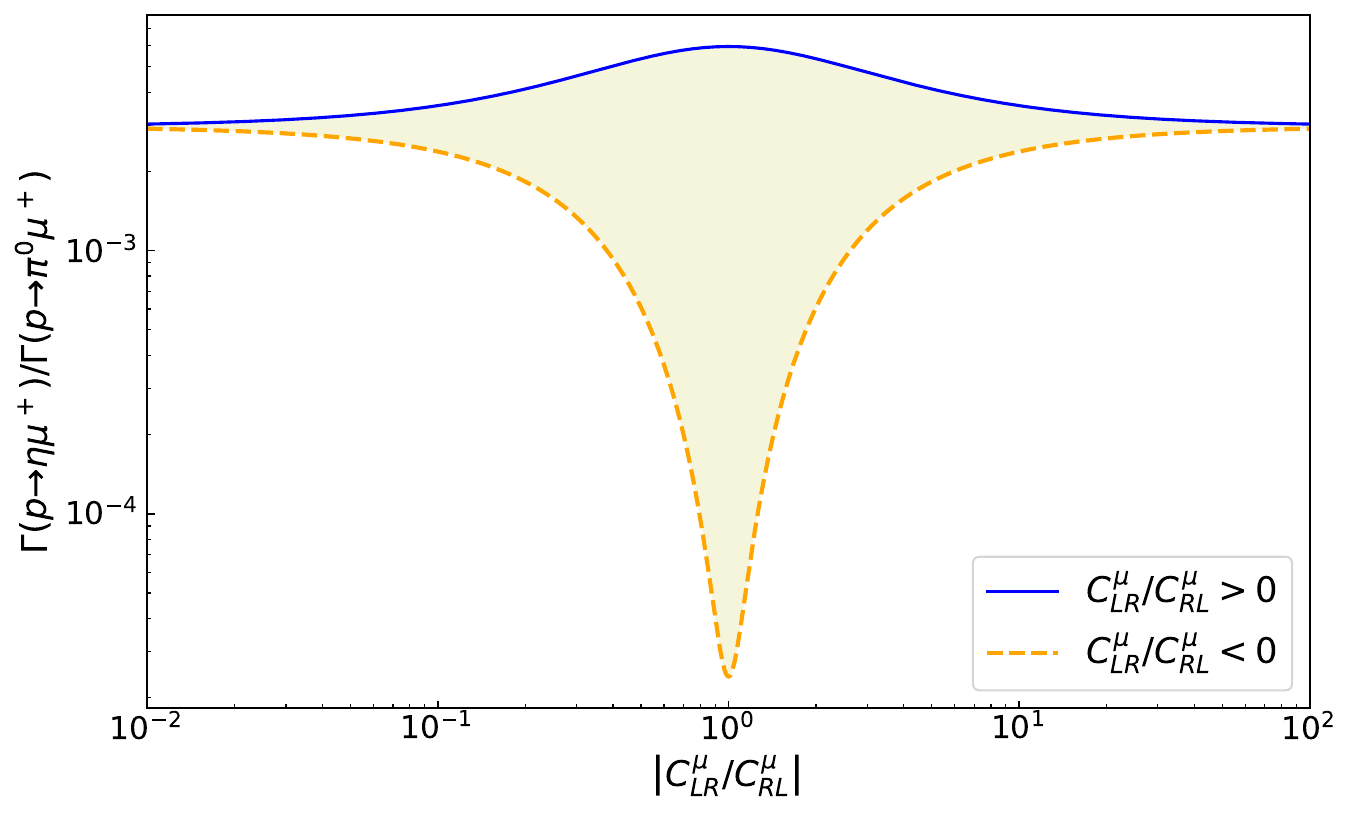}}
      \subcaptionbox{\label{fig:mupure}
      Pure only 
      }
      { 
      \includegraphics[width=0.48\textwidth]{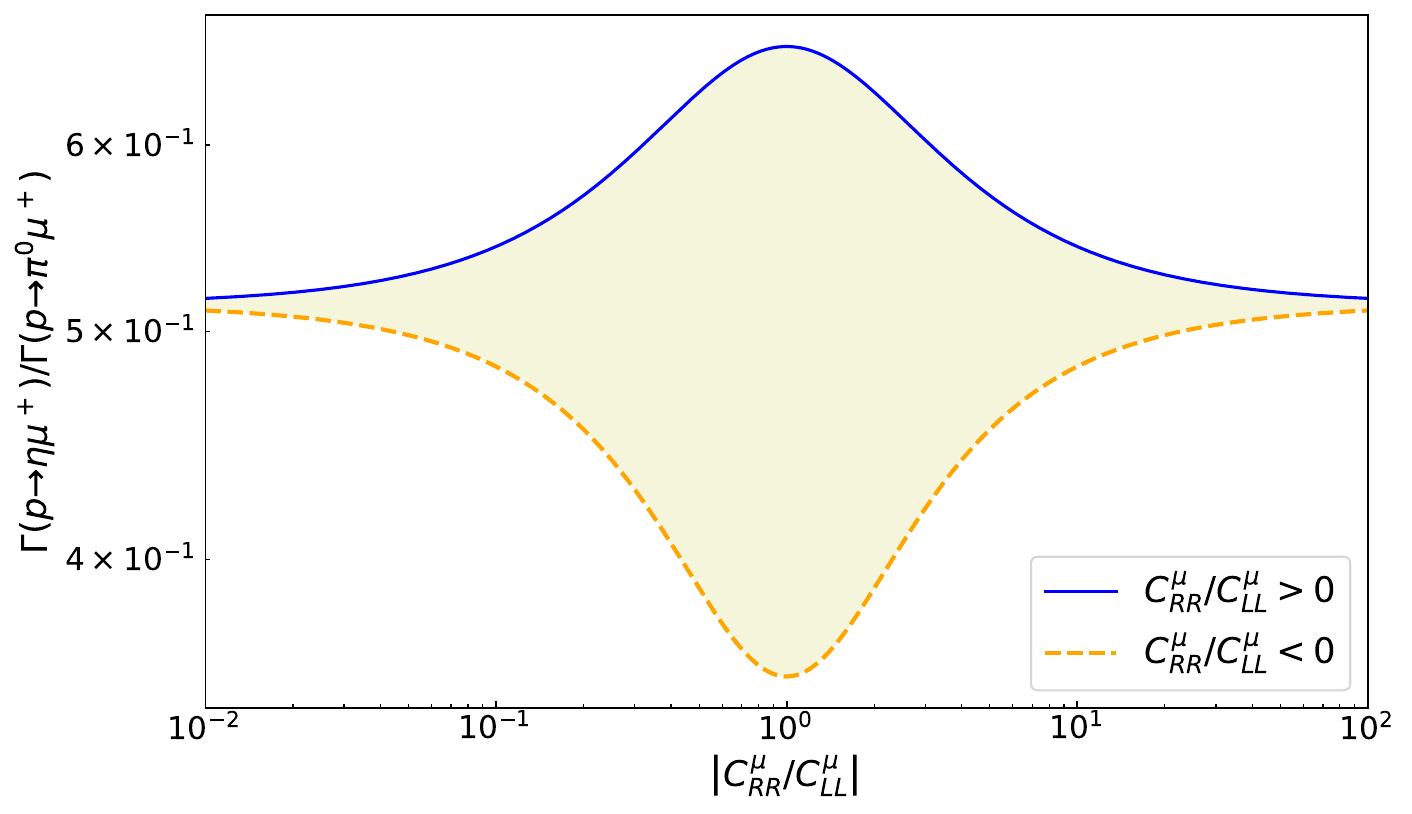}}
      \caption{The ratio $\Gamma (p \to \eta \mu^+)/\Gamma (p \to \pi^0 \mu^+)$ for the mixed-only and pure-only cases as functions of $|C_{LR}^\mu/C_{RL}^\mu|$ and $|C_{RR}^\mu/C_{LL}^\mu|$, respectively. The blue solid (orange dashed) line represents the case where the ratio of the Wilson coefficient is positive (negative), while the beige shade corresponds to the complex values of the ratio. 
      \label{fig:antimu}}
      \end{figure}  

In Fig.~\ref{fig:antimu}, we plot the ratio $\Gamma (p \to \eta \mu^+)/\Gamma (p \to \pi^0 \mu^+)$ for the mixed-only (Fig.~\ref{fig:mumix}) and pure-only (Fig.~\ref{fig:mupure}) cases as functions of $|C_{LR}^\mu/C_{RL}^\mu|$ and $|C_{RR}^\mu/C_{LL}^\mu|$, respectively. The blue solid (orange dashed) line represents the case where the ratio of the Wilson coefficient is positive (negative), while the beige shade corresponds to the complex values of the ratio. In the mixed-only case, the ratio can vary by more than an $\mathcal{O}(1)$ factor when $|C_{LR}^\mu| \simeq |C_{RL}^\mu|$, though its overall magnitude remains small due to the small value of $W^{LR}_{p\eta \ell, 0}$. The ratio converges to $\simeq 0.003$ in the limit where $|C_{LR}^\mu/C_{RL}^\mu| \gg 1$ or $\ll 1$, which is slightly larger than the value given in Eq.~\eqref{eq:poslim}. This is because in this case the expansion in terms of $m_\mu / m_p$ made in Eq.~\eqref{eqp:ppimuappr} does not work well as $|W^{LR}_{p\eta \ell, 1}| \gg |W^{LR}_{p\eta \ell, 0}|$. In contrast, for the pure-only case, the ratio can change by at most $\mathcal{O}(10)$\%, but since its magnitude is relatively large, the observational prospects are more favorable in this case compared to the mixed-only case. This result indicates that to observe this effect, it is essential to control the error in the computation of the relevant form factors to below 10\%.

\subsection{Neutrino channels}

Finally, we consider the neutrino channels. Figure~\ref{fig:matching} shows that only the operators $\mathcal{O}_{ijkl}^{(1)}$ and $\mathcal{O}_{ijkl}^{(3)}$ can induce these channels. Again, we start with examining the case where only either the mixed-type or pure-type operators are present. For the neutrino channels, the ratio is found to be 
\begin{align}
  \frac{\Gamma (n \to \eta \bar{\nu})}{\Gamma (n \to \pi^0  \bar{\nu})} = \frac{(1-m_\eta^2/m_n^2)^2}{(1-m_\pi^2/m_n^2)^2} \cdot \frac{|W_{n\eta \nu}^{\chi\chi^\prime}|^2}{|W_{n\pi^0 \nu}^{\chi\chi^\prime}|^2}
   \simeq \frac{\Gamma (p \to \eta e^+)}{\Gamma (p \to \pi^0  e^+)} 
  ~,
  \label{eq:nulim}
\end{align}
where we have used Eq.~\eqref{eq:wppinpirel} and Eq.~\eqref{eq:wetarel}. Therefore, as in the case of the positron channels, the mixed-only and pure-only cases can be distinguished through the measurement of this ratio. 

In the presence of both types of operators, the decay widths are expressed by means of the parameters in Eq.~\eqref{eq:ffind} as 
\begin{align}
  \Gamma (n \to \pi^0 \bar{\nu}) &= \frac{m_n}{32\pi} \biggl(1-\frac{m_\pi^2}{m_n^2}\biggr)^2 \frac{(1+D+F)^2 \alpha^2}{2f^2} \sum_i \left|C_{RL}^{\nu_i} - C_{LL}^{\nu_i}\right|^2~, \\ 
  \Gamma (n \to \eta \bar{\nu}) &= \frac{m_n}{32\pi} \biggl(1-\frac{m_\eta^2}{m_n^2}\biggr)^2 \frac{\alpha^2}{6 f^2} \sum_i
  \left|(1+D-3F)C_{RL}^{\nu_i} +(3-D+3F) C_{LL}^{\nu_i}\right|^2 
   ~.
\end{align}
These decay widths have a similar form to those for the positron channels but depend on a larger number of Wilson coefficients since all neutrino generations participate in these processes. This complexity makes it difficult to extract information about the Wilson coefficients from the ratio $\Gamma (n \to \pi^0 \bar{\nu})/\Gamma (n \to \eta \bar{\nu})$. The same challenge applies to the ratios between neutrino and anti-lepton channels, such as $\Gamma (n \to \pi^0 \bar{\nu})/\Gamma(p\to \pi^0 \ell^+)$. However, these ratios become powerful probes when considering specific ultraviolet (UV) models, as we will see in the next section.

\section{Specific examples}
\label{sec:models}

Now, we demonstrate that the following ratios of the partial decay widths are sensitive to the structure of the underlying theory by considering specific UV models: 
\begin{itemize}
  \item $\Gamma (p \to \eta \ell^+)/ \Gamma (p \to \pi^0 \ell^+)$
  \item $\Gamma (n \to \eta \bar{\nu})/ \Gamma (n\to \pi^0 \bar{\nu})$
  \item $\Gamma (n\to \pi^0 \bar{\nu})/\Gamma (p \to \pi^0 \ell^+)$
\end{itemize}

\subsection{Minimal SUSY SU(5) with high-scale SUSY}
\label{sec:minimalsu5}

\begin{figure}
  \centering
  \includegraphics[height=70mm]{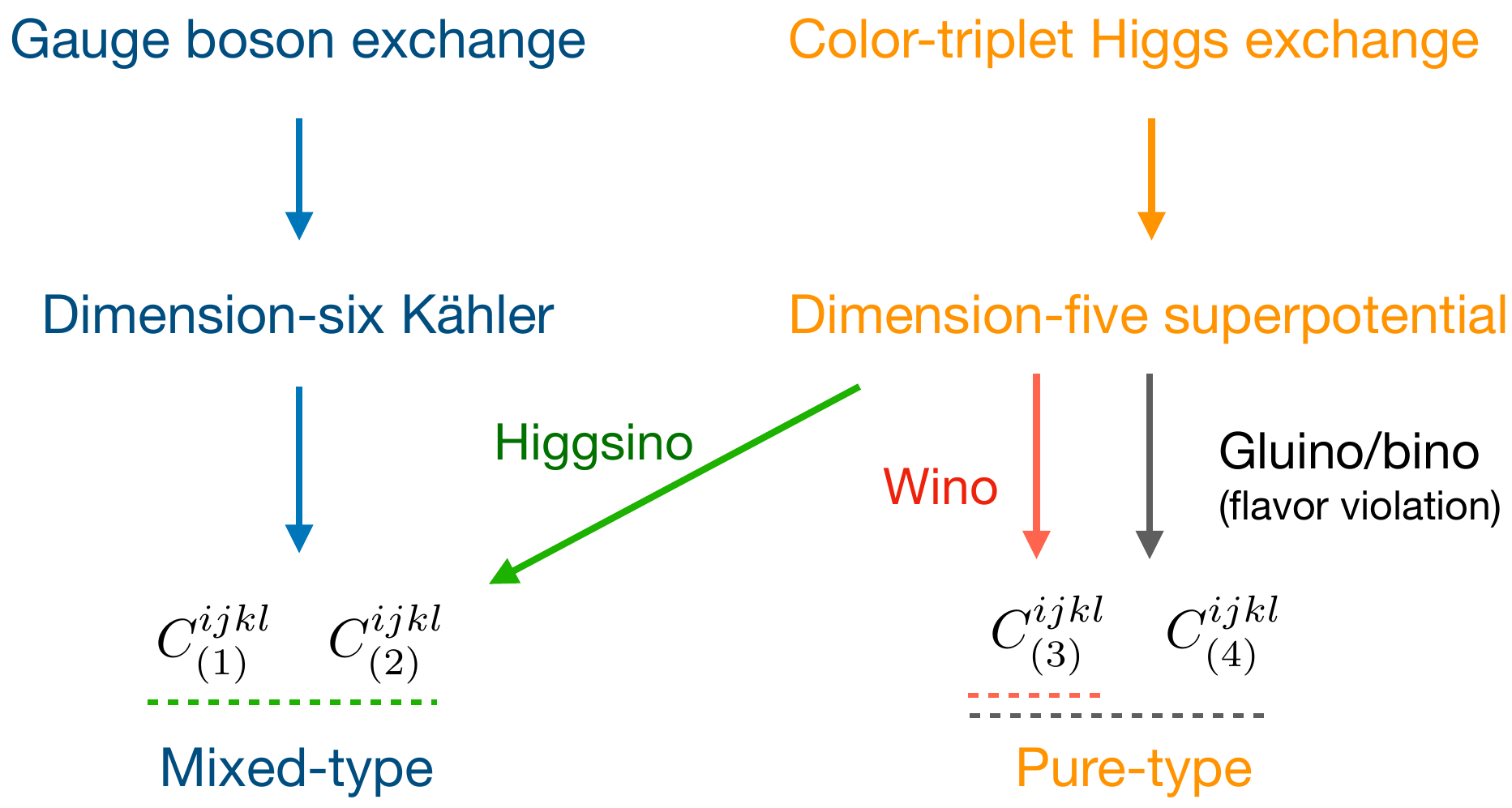}
  \caption{Each contribution to the operators in Eq.~\eqref{eq:fourfermidef} in the minimal supersymmetric SU(5) GUT.
  }  
  \label{fig:susy}
\end{figure}

As a specific UV model, we first consider the minimal SUSY SU(5) GUT~\cite{Dimopoulos:1981zb, Sakai:1981gr}. In this model, nucleon decay is induced by the exchange of color-triplet Higgs fields and SU(5) gauge bosons, which generate dimension-five superpotential operators and dimension-six K\"{a}hler-potential operators~\cite{Weinberg:1981wj, Sakai:1981pk}, respectively. The dimension-five operators exhibit a pure chirality structure, while the dimension-six ones are of the mixed type. At the SUSY mass threshold, the dimension-six operators are matched onto the mixed-type operators $\mathcal{O}^{(1)}_{ijkl}$ and $\mathcal{O}^{(2)}_{ijkl}$ at tree level. On the other hand, the dimension-five operators are matched to the operators in Eq.~\eqref{eq:fourfermidef} via gaugino/higgsino exchange one-loop diagrams. In the absence of flavor violation in the sfermion sector, the higgsino and wino exchange processes dominate. The higgsino exchange contribution generates mixed-type operators $\mathcal{O}^{(1)}_{ijkl}$ and $\mathcal{O}^{(2)}_{ijkl}$ due to chirality flips in the Yukawa interactions. On the other hand, the wino exchange contribution conserves chirality, leading to pure-type operators. Since the wino does not interact with right-handed fermions, this process contributes only to $\mathcal{O}^{(3)}_{ijkl}$. These connections are summarized in Fig.~\ref{fig:susy}. For details on the calculation of nucleon decay in the minimal SUSY SU(5), see Refs~\cite{Goto:1998qg, Nagata:2013sba, Nagata:2013ive}.

We study the effects of these contributions on the ratios of nucleon decay widths. To this end, we first note that the gauge-boson exchange contribution, the dimension-five higgsino exchange contribution, and the dimension-five wino exchange contribution are proportional to $M_X^{-2}$, $\mu_H/(M_{H_C} M^2_{\mathrm{SUSY}})$, and $M_2/(M_{H_C} M^2_{\mathrm{SUSY}})$, respectively. Here, $M_X$, $M_{H_C}$, $\mu_H$, $M_2$, and $M_{\mathrm{SUSY}}$ are the masses of the SU(5) gauge bosons, color-triplet Higgs fields, higgsino, wino, and SUSY scalar fields, respectively, 
and we assumed $M_{\mathrm{SUSY}}\gtrsim |M_2|,|\mu|$.
Thus, we can enhance the gauge-boson exchange contribution relative to the color-triplet Higgs exchange contribution by decreasing $M_X$ while keeping the other masses fixed, or vice versa. Similarly, the higgsino (wino) exchange contribution becomes dominant when $|\mu_H| \gg |M_2|$ ($|\mu_H| \ll |M_2|$).

It is well-known that if SUSY particle masses are at the TeV scale, the contribution from dimension-five operators becomes too large to avoid the current experimental limits on the $p \to K^+ \bar{\nu}$ decay channel~\cite{Goto:1998qg, Murayama:2001ur}. The aim of the following analysis is to illustrate how the ratio of nucleon decay widths depends on the model parameters. To facilitate this analysis without being constrained by current experimental limits, we assume $M_{\mathrm{SUSY}}$ to be at the multi-TeV scale, for which the experimental limits are found to be evaded~\cite{Hisano:2013exa, Nagata:2013sba, Nagata:2013ive, Evans:2015bxa, Ellis:2015rya, Ellis:2016tjc, Ellis:2019fwf, Evans:2019oyw, Hamaguchi:2022yjo}. This assumption is also beneficial because it allows the observed Higgs boson mass to be naturally explained within the framework of the minimal supersymmetric Standard Model (MSSM). In the following analysis, we adopt a mini-split type mass spectrum~\cite{Hall:2011jd, Hall:2012zp, Ibe:2011aa, Ibe:2012hu, Arvanitaki:2012ps, Arkani-Hamed:2012fhg, Evans:2013lpa, Evans:2013dza} for SUSY particles, where fermionic SUSY particles can have masses lower than $M_{\mathrm{SUSY}}$ by orders of magnitude.
This scenario can naturally occur if fermion masses are generated through higher-order radiative corrections, such as those found in anomaly mediation~\cite{Randall:1998uk, Giudice:1998xp}. For other potential contributions to gaugino masses, see, \textit{e.g.}, Refs.~\cite{Pierce:1996zz, Pomarol:1999ie, Nelson:2002sa, Hsieh:2006ig, Gupta:2012gu, Nakayama:2013uta, Harigaya:2013asa, Evans:2014xpa, Evans:2019oyw}. It is known that for this type of SUSY mass spectrum, gauge coupling unification is still achieved with good accuracy~\cite{Hisano:2013cqa}.

\begin{figure}
  \centering
  \subcaptionbox{\label{fig:ratioMHcmuH1TeVfinal}
  $\mu_H = 1~\mathrm{TeV}$
  }
  {\includegraphics[height=80mm]{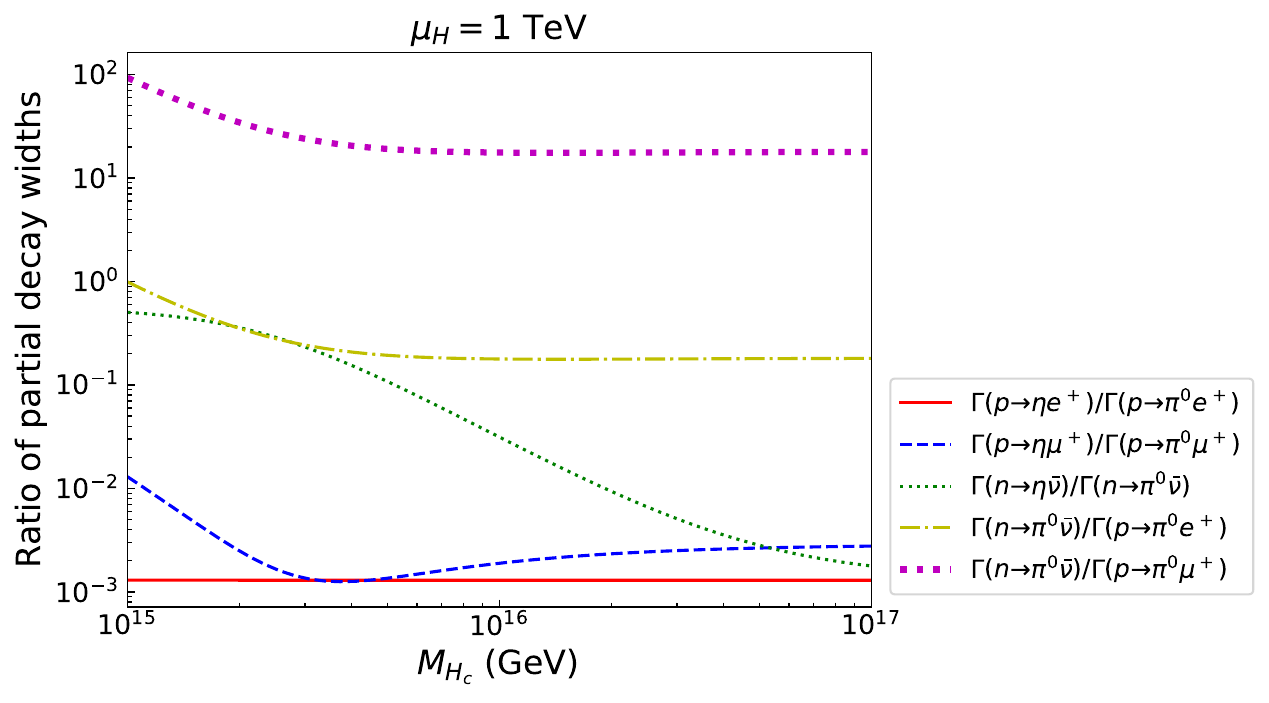}}\\\vspace{.5cm}
  \subcaptionbox{\label{fig:ratioMHcmuH=MSUSYfinal}
  $\mu_H = M_{\mathrm{SUSY}}$
  }
  { 
  \includegraphics[height=80mm]{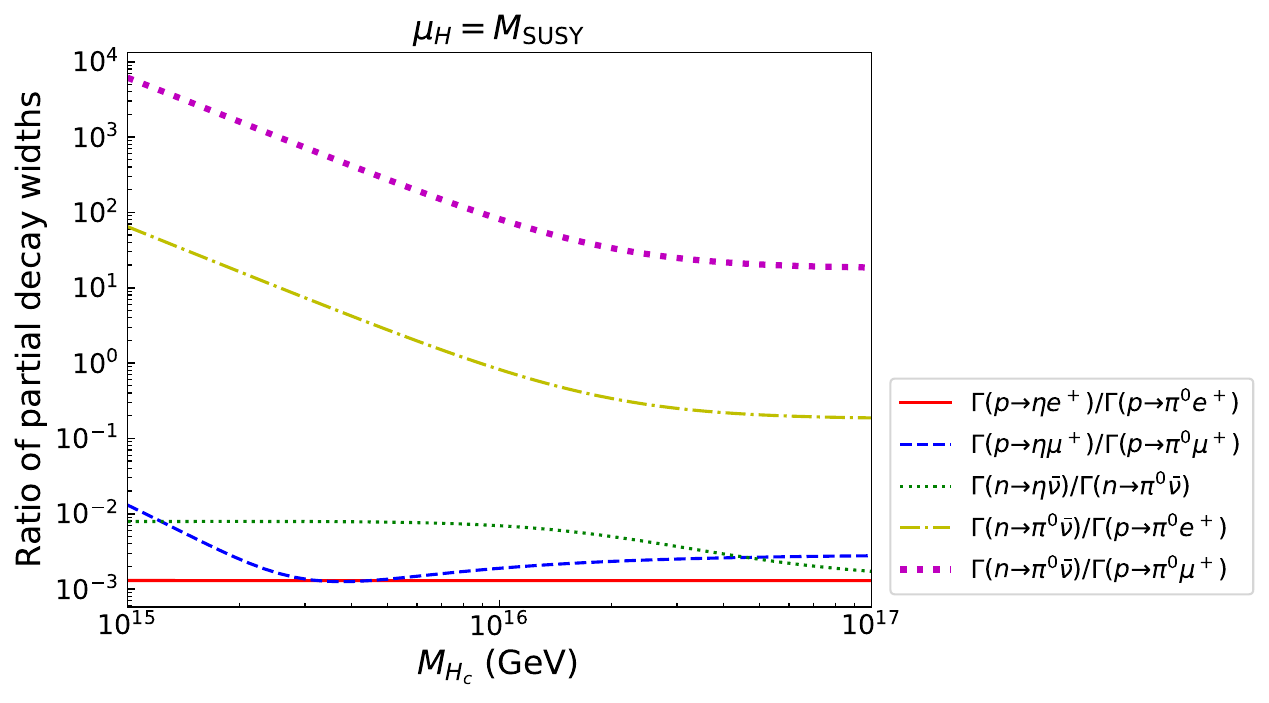}}
  \caption{Ratios of the decay widths as functions of the color-triplet mass $M_{H_C}$, where we take $M_X = 10^{16}~\mathrm{GeV}$, $M_2 = 1~\mathrm{TeV}$, $M_3 = 10~\mathrm{TeV}$, $M_{\mathrm{SUSY}} = 100~\mathrm{TeV}$, and $\tan \beta = 3$. 
  \label{fig:ratioMHc}
  }
  \end{figure}  

In Fig.~\ref{fig:ratioMHc}, we plot the ratios of the decay widths as functions of the color-triplet mass $M_{H_C}$, where we take $M_X = 10^{16}~\mathrm{GeV}$, $M_2 = 1~\mathrm{TeV}$, $M_3 = 10~\mathrm{TeV}$, $M_{\mathrm{SUSY}} = 100~\mathrm{TeV}$, and $\tan \beta = 3$,
with $\tan \beta \equiv \langle H_u^0 \rangle / \langle H_d^0 \rangle $ the ratio of the vacuum expectation values (VEVs) of the two Higgs doublets in the MSSM.\footnote{In the numerical calculation, we run the gauge and Yukawa couplings from low energy to the GUT scale, 
defined by the scale where the U(1)$_Y$ and SU(2)$_L$ couplings unify, via two and one loop RGEs, respectively, with threshold matching at $M_{\mathrm{gaugino}}=\sqrt{M_2 M_3}$ and $M_{\mathrm{ SUSY}}$. The dimension-five and -six operators are set at the GUT scale with the GUT CP phases set to be zero, for simplicity. The Wilson coefficients are then evolved down to hadronic scale, with the matching at the $M_{\mathrm{ SUSY}}$ for the dimension five operators. We input the SUSY parameters at the SUSY mass threshold, not at the GUT scale.
}
The higgsino mass is taken to be $\mu_H = 1~\mathrm{TeV}$ and $\mu_H = M_{\mathrm{SUSY}}$ in Fig.~\ref{fig:ratioMHcmuH1TeVfinal} and Fig.~\ref{fig:ratioMHcmuH=MSUSYfinal}, respectively. These plots show that the ratio $\Gamma (p \to \eta e^+)/\Gamma(p \to \pi^0 e^+)$, represented by the red solid lines, scarcely depends on $M_{H_C}$, indicating that these decay channels are induced dominantly by the gauge-boson exchange. This follows from a generic feature of the color-triplet Higgs exchange process, where the interactions with the first-generation fermions tend to be suppressed by small Yukawa couplings. On the other hand, the ratio $\Gamma (p \to \eta \mu^+)/\Gamma(p \to \pi^0 \mu^+)$ shown in the blue dashed lines increases as $M_{H_C}$ decreases, with little dependence on the choice of $\mu_H$; we thus conclude that this enhancement is attributed to the dimension-five wino-exchange processes. The higgsino-exchange contribution to the effective operators~\eqref{eq:fourfermidef} with the first- and second-generation fermions is additionally suppressed by small Yukawa couplings, resulting in extremely small contributions to $p \to \pi^0 \ell^+$ and $p \to \eta \ell^+$. The effect of $M_{H_C}$ can also be seen in the mixed channels, $\Gamma (n \to \pi^0 \bar{\nu})/\Gamma (p \to \pi^0 \ell^+)$ (the yellow dash-dotted line for $\ell = e^+$ and purple dotted line for $\ell = \mu$). In these cases, the results depend also on the choice of the higgsino mass, as the third-generation neutrino can contribute to the neutrino channels. As for $\Gamma (n \to \eta \bar{\nu})/\Gamma (n \to \pi^0 \bar{\nu})$, depicted by the green dotted lines, its $M_{H_C}$-dependence is considerable for $\mu_H = 1~\mathrm{TeV}$ but less significant for $\mu_H = M_{\mathrm{SUSY}}$. This behavior can be understood from Eq.~\eqref{eq:poslim} and Eq.~\eqref{eq:nulim}. For $\mu_H = M_{\mathrm{SUSY}}$, the dominant contribution comes from the gauge-boson exchange processes or the higgsino-exchange processes, both of which give rise to the mixed-type operators as shown in Fig.~\ref{fig:susy}; this results in a small value of $\Gamma (n \to \eta \bar{\nu})/\Gamma (n \to \pi^0 \bar{\nu})$. For $\mu_H = 1~\mathrm{TeV}$, on the other hand, the wino contribution is dominant for small values of $M_{H_C}$, leading to large contribution to the pure-type operators and thus to a larger value of $\Gamma (n \to \eta \bar{\nu})/\Gamma (n \to \pi^0 \bar{\nu})$.

\begin{figure}
  \centering
  \includegraphics[height=80mm]{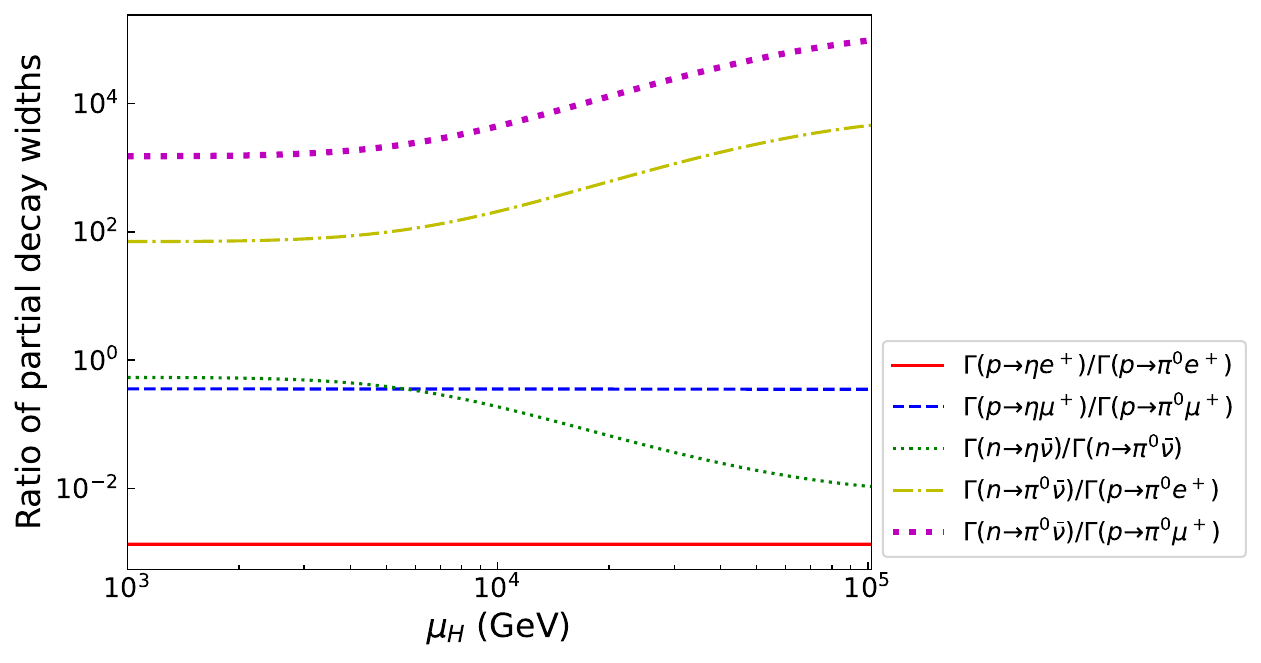}
  \caption{Ratios of the decay widths as functions of the higgsino mass $\mu_H$ for $M_X = 10^{17}~\mathrm{GeV}$, $M_{H_C} = 10^{16}~\mathrm{GeV}$, $M_2 = 1~\mathrm{TeV}$, $M_3 = 10~\mathrm{TeV}$, $M_{\mathrm{SUSY}} = 100~\mathrm{TeV}$, and $\tan \beta = 3$. 
  }  
  \label{fig:ratiomuHfinal}
\end{figure}

To study the interplay between the higgsino and wino contributions in more detail, in Fig.~\ref{fig:ratiomuHfinal}, we plot the ratios as functions of the higgsino mass $\mu_H$ for a larger value of the gauge-boson mass, $M_X = 10^{17}~\mathrm{GeV}$, aiming at suppressing the contribution of the dimension-six K\"{a}hler-type operators. The other parameters are taken to be $M_{H_C} = 10^{16}~\mathrm{GeV}$, $M_2 = 1~\mathrm{TeV}$, $M_3 = 10~\mathrm{TeV}$, $M_{\mathrm{SUSY}} = 100~\mathrm{TeV}$, and $\tan \beta = 3$. It is found that the ratios $\Gamma (p \to \eta \ell^+)/\Gamma(p \to \pi^0 \ell^+)$ rarely depend on the higgsino mass, namely, the higgsino contribution to  $p \to \eta \ell^+$ and $p \to \pi^0 \ell^+$ is negligible, as we anticipated above. For the anti-muon mode, the value of the ratio is close to the one for the pure-only case in Eq.~\eqref{eq:poslim}, indicating that the wino-exchange contribution is dominant. For the positron mode, however, it agrees to the mixed-only case in Eq.~\eqref{eq:poslim}, showing that the gauge-boson exchange contribution is still dominant even if we take the gauge-boson mass to be as large as $10^{17}$~GeV. Taking the results in Fig.~\ref{fig:ratioMHc} into account, we conclude that the ratio $\Gamma (p \to \eta \mu^+)/\Gamma(p \to \pi^0 \mu^+)$ is most useful to distinguish the dimension-five wino exchange and dimension-six gauge-boson exchange contributions, as it does not suffer from the contamination of the higgsino contribution. The other ratios depend on the higgsino mass, and thus can in principle be used to distinguish the wino and higgsino contributions. However, very large values of $\Gamma (n \to \pi^0 \bar{\nu})/\Gamma (p \to \pi^0 \ell^+)$ imply that it is difficult to detect both $n \to \pi^0 \bar{\nu}$ and $p \to \pi^0 \ell^+$ in future experiments. We, therefore, conclude that $\Gamma (n \to \eta \bar{\nu})/\Gamma (n \to \pi^0 \bar{\nu})$ is most useful to observe the effect of the higgsino exchange contribution. 

We note that in most of the parameter ranges depicted in Figs.~\ref{fig:ratioMHc} and \ref{fig:ratiomuHfinal}, the lifetimes of the decay channels are predicted to be beyond the sensitivity of upcoming nucleon-decay experiments. This outcome primarily stems from our illustrative choice of a large $M_X$ and $M_{\mathrm{SUSY}}$. However, reducing these parameters could bring some of the lifetimes within the detectable range of future experiments.

\subsection{Minimal SUSY SU(5) with sfermion flavor violation}
\label{sec:sfermionfv}

In the previous example, the mixed-type operators $\mathcal{O}^{(1)}_{ijkl}$ and $\mathcal{O}^{(2)}_{ijkl}$ receive multiple contributions, including those from gauge bosons and higgsinos. In contrast, for the pure-type operators, only a single pure-type operator, $\mathcal{O}^{(3)}_{ijkl}$, is induced by a sole contribution---the wino exchange process. We now consider a different scenario, still within the context of a mini-split SUSY spectrum in the MSSM, where the pure-type operators receive multiple contributions. By introducing flavor violation in the sfermion mass matrices, the bino and gluino can also induce pure-type operators, as shown in Fig.~\ref{fig:susy}. We assume the sfermion soft mass matrices to have the form 
\begin{equation}
  \widetilde{m}_{\tilde{f}}^2 = M_{\mathrm{SUSY}}^2 
  \begin{pmatrix}
    1 & \delta^{\tilde{f}}_{12} &  \delta^{\tilde{f}}_{13} \\ 
    \delta^{\tilde{f}*}_{12} & 1 &  \delta^{\tilde{f}}_{23} \\ 
    \delta^{\tilde{f}*}_{13} &  \delta^{\tilde{f}*}_{23} & 1
  \end{pmatrix}
  ~,
\end{equation}
with $\tilde{f} = \tilde{Q}_L, \tilde{u}_R, \tilde{d}_R, \tilde{L}_L, \tilde{e}_R$. For simplicity, the parameters on the right-hand side are set at the SUSY scale. In a more realistic scenario, however, these parameters could be defined at a higher energy scale, such as the GUT scale. In that case, we would expect significant RGE effects on the sfermion mass matrices at low energies. Additionally, there may be correlations among the input parameters for squark and slepton masses, as they could belong to the same GUT multiplets. In our analysis, we do not explicitly impose such correlations for illustrative purposes.

\begin{figure}
  \centering
  \subcaptionbox{\label{fig:ratiodelta13}
  $\delta^{\tilde{Q}_L}_{13}$
  }
  {\includegraphics[height=80mm]{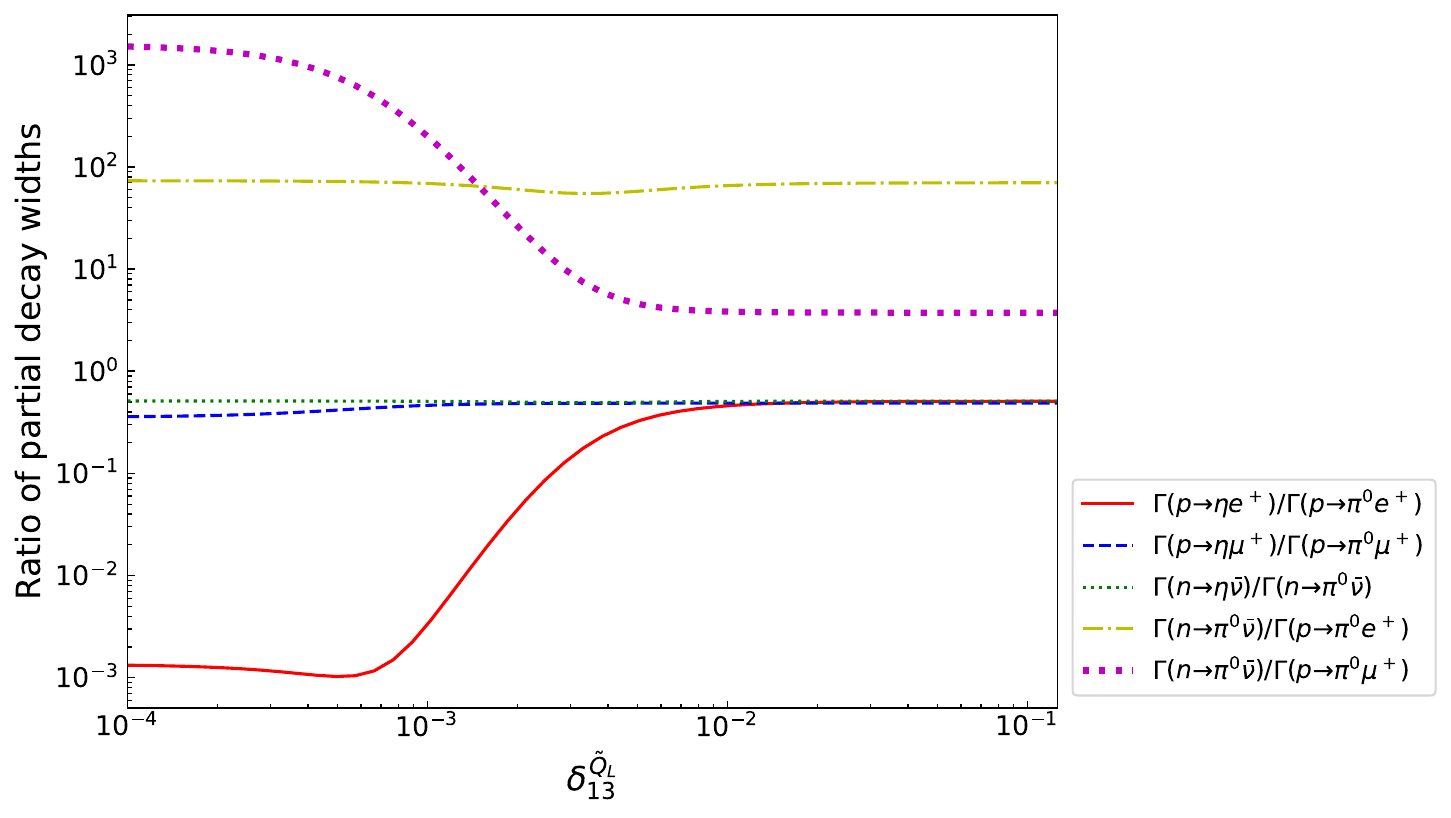}}\\\vspace{.5cm}
  \subcaptionbox{\label{fig:ratiodelta13uR}
  $\delta^{\tilde{u}_R}_{13}$
  }
  { 
  \includegraphics[height=80mm]{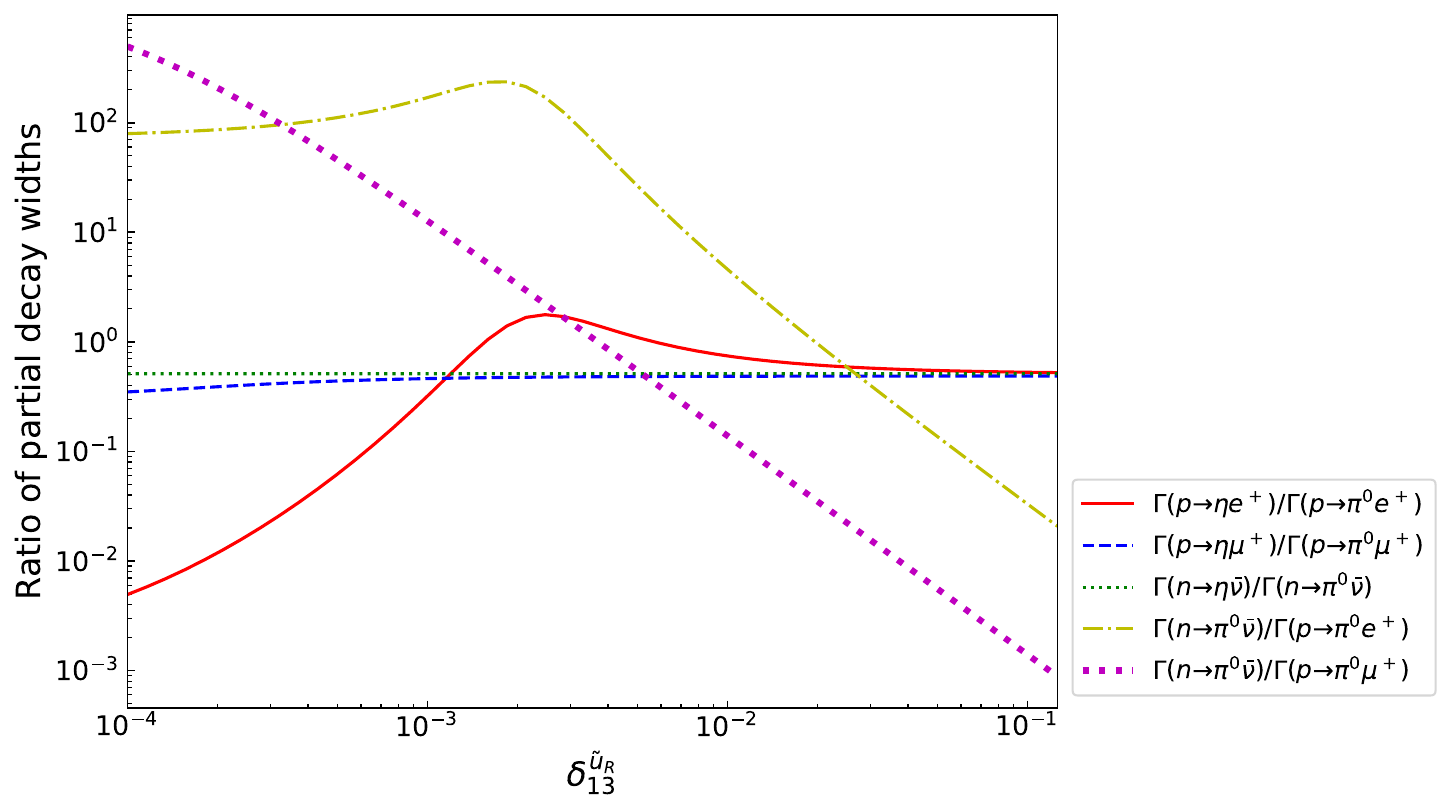}}
  \caption{Ratios of the decay widths as functions of $\delta^{\tilde{f}}_{13} $ for $M_{H_C} =10^{16}$~GeV, $M_X =  10^{17}$~GeV, $\mu_H = 200$~GeV,  $M_1 = 5$~TeV, $M_2 = 1$~TeV, $M_3 = 10$~TeV, $M_{\mathrm{SUSY}} = 100$~TeV, and $\tan \beta = 3$.
  \label{fig:ratiofv}
  }
  \end{figure}  

In Fig.~\ref{fig:ratiofv}, we show the ratios of the decay widths as functions of $\delta^{\tilde{f}}_{13}$. As we aim at examining the gaugino contributions to the pure-type operators, we take $M_X = 10^{17}~\mathrm{GeV}$ to suppress the gauge boson contribution and $\mu_H = 200$~GeV to suppress the higgsino contribution. The other parameters are set to be  $M_{H_C} =10^{16}$~GeV, $M_1=5$~TeV,
$M_2 = 1$~TeV, $M_3 = 10$~TeV, $M_{\mathrm{SUSY}} = 100$~TeV, and $\tan \beta = 3$. We use the matching conditions at $M_{\mathrm{SUSY}}$ given in Ref.~\cite{Nagata:2013sba,Nagata:2013ive}.

Figure~\ref{fig:ratiodelta13} is for $\tilde{f} = \tilde{Q}_L$. In this case, the gluino and bino exchange processes induce only $\mathcal{O}^{(3)}_{ijkl}$, similarly to the wino exchange contribution. We see that the new contribution significantly enhances the ratio $\Gamma (p \to \eta e^{+})/\Gamma (p \to \pi^0 e^+)$, and once it dominates the other contributions, $\Gamma (p \to \eta \ell^+)/\Gamma(p \to \pi^0 \ell^+)$ and $\Gamma (n \to \eta \bar{\nu})/\Gamma (n \to \pi^0 \bar{\nu})$ lead to the values for the pure-only case in Eq.~\eqref{eq:poslim} and Eq.~\eqref{eq:nulim}. We also find that the gluino/bino contribution scarcely changes $\Gamma (n \to \pi^0 \bar{\nu})/\Gamma (p \to \pi^0 e^+)$. On the other hand, there is considerable change in $\Gamma (n \to \pi^0 \bar{\nu})/\Gamma (p \to \pi^0 \mu^+)$, making this ratio a useful probe of the flavor violation in $\tilde{Q}_L$.  

On the other hand, for $\tilde{f} = \tilde{u}_R$, the gluino and bino contributions induce only $\mathcal{O}^{(4)}_{ijkl}$. In this case, the behavior of $\Gamma (n \to \pi^0 \bar{\nu})/\Gamma (p \to \pi^0 \ell^+)$ gets drastically different from that for $\tilde{f} = \tilde{Q}_L$, as shown in Fig.~\ref{fig:ratiodelta13uR}. This is because $\mathcal{O}^{(4)}_{ijkl}$ contributes to $p \to \pi^0 \ell^+$ but not to $n \to \pi^0 \bar{\nu}$, so the ratios keep decreasing as $\delta^{\tilde{u}_R}_{13}$ gets larger. The other ratios lead to the values for the pure-only case in Eq.~\eqref{eq:poslim} and Eq.~\eqref{eq:nulim} for large values of $\delta^{\tilde{u}_R}_{13}$ as in the previous case. Therefore, the ratios $\Gamma (n \to \pi^0 \bar{\nu})/\Gamma (p \to \pi^0 \ell^+)$ play a crucial role in discriminating this case from others. 

Finally, we note that for the parameter choices in Fig.~\ref{fig:ratiofv}, the nucleon decay lifetimes could potentially be within the detection range of upcoming experiments or even ruled out by current experimental limits for large values of $\delta^{\tilde{f}}_{13}$. However, these limits can be easily avoided if $M_{\mathrm{SUSY}}$ or $M_{H_C}$ is increased.

\section{Conclusion and discussion}
\label{sec:conclusion}

We have investigated the ratios of partial decay widths in strangeness-conserving nucleon decay processes, aiming to explore the chirality structure of the underlying theory that generates baryon-number violating operators. In Sec.~\ref{sec:chirality}, we present a general analysis using an effective field theory approach, finding that the ratio $\Gamma (p \to \eta \ell^+)/\Gamma (p \to \pi^0 \ell^+)$ can differentiate between pure-type and mixed-type operators. Ratios involving neutrinos, such as $\Gamma (n \to \pi^0 \bar{\nu})/\Gamma (n \to \eta \bar{\nu})$ and $\Gamma (n \to \pi^0 \bar{\nu})/\Gamma(p\to \pi^0 \ell^+)$, also contain information about the chirality structure. However, extracting this information is challenging due to the contribution of all three neutrino generations. Nevertheless, these channels can serve as powerful probes when examining specific UV models. We illustrate these points in Sec.~\ref{sec:models} using the minimal SUSY SU(5) GUT model with a mini-split SUSY spectrum. As discussed in Sec.~\ref{sec:minimalsu5}, the ratios $\Gamma (p \to \eta \mu^+)/\Gamma(p \to \pi^0 \mu^+)$ and $\Gamma (n \to \eta \bar{\nu})/\Gamma (n \to \pi^0 \bar{\nu})$ can distinguish between pure-type contributions, induced by wino exchange processes, and mixed-type contributions, induced by higgsino/gauge-boson exchange. Additionally, in Sec.~\ref{sec:sfermionfv}, we consider the scenario where flavor violation occurs in the sfermion mass matrices. In this case, gluino and bino exchange processes provide additional contributions to the pure-type operators, and we find that the ratios $\Gamma (n \to \pi^0 \bar{\nu})/\Gamma (p \to \pi^0 \ell^+)$ are particularly sensitive to the presence of such flavor violation.

As we have seen in our work, measuring the $n \to \eta \bar{\nu}$ decay channel could be crucial for probing the structure of the underlying model. However, the most recent limit on this decay was imposed by an old experiment, IMB-3~\cite{McGrew:1999nd}, with no updates from the Super-Kamiokande experiment. Considering the significance of this decay channel, it is important to develop search strategies for it in current and future nucleon decay experiments.

In some cases, the model dependence of the ratio is around $\mathcal{O}(10)$\%. To observe this effect, it is essential to control the theoretical error to be within 10\%. As reviewed in Sec.~\ref{sec:ff}, the current accuracy of lattice calculations for hadron matrix elements is approximately 10\%, with the exception of the $\eta$ channels, where the omission of disconnected diagram contributions might introduce additional systematic uncertainties~\cite{Aoki:2017puj}. It is also desirable to compute $W_{p \eta \ell, 1}^{LR}$ and $W_{p \eta \ell, 1}^{LL}$ directly, rather than the specific combination in Eq.~\eqref{eq:wmudef}. We expect improvements in this area in the future. Furthermore, as discussed in Sec.~\ref{sec:cpt}, the uncertainty in the ratios of matrix elements may be significantly smaller than that in the matrix elements themselves---it would be beneficial to confirm this expectation through the direct method in future lattice calculations.

\section*{Acknowledgments}

This work was supported by JSPS KAKENHI Grant Numbers 24H02244 (KH), 24K07041 (KH), 21K13916 (NN), 22KJ1022 (SH), and 24KJ0913 (HT).



\bibliographystyle{utphysmod}
\bibliography{ref}


\end{document}